\documentclass[aps,prd, twocolumn, superscriptaddress, nofootinbib, floatfix]{revtex4-1}
\usepackage{xcolor}
\usepackage{graphicx}
\usepackage{wrapfig}
\usepackage{amsmath}
\usepackage{amsfonts}
\usepackage{amssymb}
\usepackage{multirow}
\usepackage{slashed}
\usepackage{physics}
\usepackage{float}
\usepackage{bm}
\usepackage{soul}
\usepackage[colorlinks=true, linkcolor=blue, citecolor=blue, urlcolor=blue]{hyperref}
\DeclareRobustCommand{\Eq}[1]{Eq.~\eqref{eq:#1}}

\DeclareRobustCommand{\fig}[1]{Fig.~\ref{fig:#1}}

\DeclareRobustCommand{\app}[1]{App.~\ref{app:#1}}
\DeclareRobustCommand{\sec}[1]{Sec.~\ref{sec:#1}}

\DeclareRobustCommand{\tb}[1]{Table~\ref{tb:#1}}

\DeclareRobustCommand{\refcite}[1]{Ref.~\cite{#1}}

\newcommand\bets{\begin{table*}}
\newcommand\eets[1]{\label{tb:#1}\end{table*}}

\usepackage[normalem]{ulem} 

\begin{document}

\widetext

\title{Continuum-extrapolated NNLO Valence PDF of Pion at the Physical Point}
\author{Xiang Gao}
\email{gaox@anl.gov}
\affiliation{Physics Division, Argonne National Laboratory, Lemont, IL 60439, USA}
\author{Andrew D. Hanlon}
\affiliation{Physics Department, Brookhaven National Laboratory, Upton, NY 11973, USA}
\author{Nikhil Karthik}
\affiliation{Department of Physics, College of William \& Mary, Williamsburg, VA 23185, USA}
\affiliation{Thomas Jefferson National Accelerator Facility, Newport News, VA 23606, USA}
\author{Swagato Mukherjee}
\affiliation{Physics Department, Brookhaven National Laboratory, Upton, NY 11973, USA}
\author{Peter Petreczky}
\affiliation{Physics Department, Brookhaven National Laboratory, Upton, NY 11973, USA}
\author{Philipp Scior}
\affiliation{Physics Department, Brookhaven National Laboratory, Upton, NY 11973, USA}
\author{Shuzhe Shi}
\affiliation{Department of Physics and Astronomy, Stony Brook University, Stony Brook, NY 11794, USA}
\author{Sergey Syritsyn}
\affiliation{Department of Physics and Astronomy, Stony Brook University, Stony Brook, NY 11794, USA}
\affiliation{RIKEN-BNL Research Center, Brookhaven National Lab, Upton, NY, 11973, USA}
\author{Yong Zhao}
\affiliation{Physics Division, Argonne National Laboratory, Lemont, IL 60439, USA}
\author{Kai Zhou}
\affiliation{Frankfurt Institute for Advanced Studies (FIAS), D-60438 Frankfurt am Main, Germany}

\begin{abstract}
We present lattice QCD calculations of valence parton distribution function (PDF) of pion employing next-to-next-leading-order (NNLO) perturbative QCD matching. Our calculations are based on three gauge ensembles of 2+1 flavor highly improved staggered quarks and Wilson--Clover valance quarks, corresponding to pion mass $m_\pi=140$~MeV at a lattice spacing $a=0.076$~fm and $m_\pi=300$~MeV at $a=0.04, 0.06$~fm. This enables us to present, for the first time, continuum-extrapolated lattice QCD results for NNLO valence PDF of the pion at the physical point. Applying leading-twist expansion for renormalization group invariant (RGI) ratios of bi-local pion matrix elements with NNLO Wilson coefficients we extract $2^{\mathrm{nd}}$, $4^{\mathrm{th}}$ and $6^{\mathrm{th}}$ Mellin moments of the PDF. We reconstruct the Bjorken-$x$ dependence of the NNLO PDF from real-space RGI ratios using a deep neural network (DNN) as well as from momentum-space matrix elements renormalized using a hybrid-scheme. All our results are in broad agreement with the results of global fits to the experimental data carried out by the xFitter and JAM collaborations. 
%
%
\end{abstract}

\date{\today}
\maketitle
\section{Introduction}\label{sec:intro}

Pions are the Nambu--Goldstone bosons of QCD with massless quarks due to spontaneous chiral symmetry breaking. Understanding the structure of the pion, therefore, plays a central role in the study of the strong interaction, in particular in clarifying the relation between hadron mass and hadron structure~\cite{Aguilar:2019teb}. 
The collinear partonic structure of the pion is encoded in the
parton distribution functions (PDFs), which describe the collinear-momentum fraction $x$ of a hadron carried by the partons, and can be extracted from deep-inelastic scattering experiments.
However, the study of the pion PDFs from experiment is more difficult than the study of the nucleon PDFs
due to the sparseness of the experimental data. As a result, the pion PDF is less constrained
than the unpolarized quark nucleon PDF. Therefore, lattice QCD calculations could play an important role in constraining
the pion PDF. The PDFs are defined as the Fourier transform of light-cone correlation functions~\cite{Collins:2011zzd} and, therefore, cannot be computed directly on a Euclidean lattice. The Mellin moments of the PDFs can be directly calculated on the lattice, but
in practice the calculations are limited to only the lowest moments due to decreasing signal-to-noise ratios and the power divergent operator mixing. 
In the case of the pion, the three lowest moments have been calculated \cite{Oehm:2018jvm}.

In recent years, significant progress has been made since large-momentum effective theory (LaMET)~\cite{Ji:2013dva,Ji:2014gla,Ji:2020ect} was proposed. LaMET makes it possible to calculate $x$-dependent PDFs on the lattice by computing the boosted matrix elements of equal-time extended operators; which, for a quark PDF, is the bilocal quark-bilinear operator,
\begin{align}
	O_\Gamma(z)\equiv \overline{\psi}(z)\Gamma W(z,0)\psi\,,
	\label{eq:operator}
\end{align}
where $W(z,0)$ is the Wilson line connecting the quark and antiquark fields, $\psi$ and $\bar\psi$, to preserve gauge invariance. For the unpolarized PDF, $\Gamma$ can be either $\Gamma=\gamma_z$ or $\Gamma=\gamma_0$. The choice $\Gamma=\gamma_0$ has several advantages including free of the operator mixing due to the explicit chiral symmetry breaking at finite lattice spacings~\cite{Constantinou:2017sej, Chen:2017mzz}, and the absence of additional higher-twist effects proportional to $z^\mu$~\cite{Radyushkin:2017cyf, Bhattacharya:2022aob}. The Fourier transform of these equal-time matrix elements defines the quasi-PDF (qPDF) which, for hadron states with large momentum, can be perturbatively matched to a light-cone PDF~\cite{Xiong:2013bka,Ji:2020ect} up to certain power corrections. Using the same operators, several approaches have also been developed to extract either the Mellin moments of PDFs or $x$-dependent PDFs, namely the Ioffe-time pseudo-distributions or the pseudo-PDF (pPDF)~\cite{Radyushkin:2017cyf, Orginos:2017kos}. Alternatively, other approaches have also been proposed such as the short-distance expansion of the current-current correlator~\cite{Braun:2007wv, Ma:2017pxb}, the operator product expansion (OPE) of a Compton amplitude in the unphysical region~\cite{Chambers:2017dov}, the hadronic tensor~\cite{Liu:1993cv}, the heavy-quark operator product expansion~\cite{Detmold:2005gg,Detmold:2021uru} and so on. Significant progress has been made for the study of nucleon isovector PDFs~\cite{Bhat:2022zrw,Bringewatt:2020ixn, Fan:2020nzz, Alexandrou:2020qtt, Bhat:2020ktg, Alexandrou:2019lfo,Alexandrou:2018pbm,Liu:2018uuj,Liu:2018hxv,Lin:2018qky,Chen:2018xof,Alexandrou:2019lfo,Alexandrou:2018eet,Alexandrou:2018pbm,Joo:2019jct,Joo:2020spy,Karpie:2021pap,Egerer:2021ymv} (for reviews, see Refs.~\cite{Ji:2020ect,Constantinou:2020hdm,Constantinou:2022yye}), the flavor decomposition of nucleon PDFs~\cite{Alexandrou:2021oih,Alexandrou:2020uyt}, gluon PDFs~\cite{Fan:2018dxu, Fan:2020cpa, HadStruc:2021wmh} as well as PDFs beyond leading-twist~\cite{Bhattacharya:2021moj, Bhattacharya:2020jfj,Bhattacharya:2020xlt,Bhattacharya:2020cen}.

Several lattice QCD calculations of the pion PDF have been performed recently 
\cite{Fan:2021bcr,Gao:2021hxl, Gao:2020ito, Zhang:2018nsy, Izubuchi:2019lyk, Sufian:2020vzb, Sufian:2019bol, Zhang:2018nsy, JeffersonLabAngularMomentumJAM:2022aix, Karpie:2018zaz, Karpie:2021pap, Shugert:2020tgq,Chen:2018fwa,Joo:2019bzr} with pion masses heavier than the physical point.  There have also been exploratory lattice studies of the structure of the pion radial excitation~\cite{Gao:2021hvs}, and of the pion in QCD-like theories~\cite{Karthik:2021qwz,Karthik:2022fdb}. In all of these studies, the next-to-leading order (NLO) perturbative matching between the Euclidean time quantities
and the light cone PDF have been used. Recently, next-to-next-to-leading order (NNLO) perturbative matching has been computed~\cite{Li:2020xml,Chen:2020ody}, which is supposed to be more reliable by reducing the perturbation theory uncertainty.
Very recently the NNLO matching has been used to study the pion PDF using the so- called hydbrid
renormalization scheme in $x$-space~\cite{Gao:2021dbh}. Additionally, an NNLO matching 
in real-space was recently performed in the case of the unpolarized proton PDF
in Ref.~\cite{Bhat:2022zrw}. The goal of this paper is to study the valence PDF of the pion for physical
quark masses with both real-space and $x$-space matching at NNLO. The lattice
calculations are performed at a single lattice spacing, $a=0.076$ fm. Combining this calculation
with the previous ones on finer lattices, but with larger than physical quark mass, we provide
estimates of the valence pion PDF and its moments in the continuum limit at the physical point. 

The rest of the paper is organized as follows. In \sec{latset} we present our lattice setup. In \sec{c2pt} we discuss the lattice calculations of the pion two point functions, while in \sec{c3pt} we discuss the extraction of the matrix elements of the quasi-PDF operator. In \sec{matching} we briefly review the ratio scheme renormalization used in this work. In \sec{moments} we discuss the determination of even moments of the valence pion PDF using short distance factorization. In \sec{modelfit} we discuss the determination of the valence pion PDF through a model dependent fit of
the lattice data. A determination of the valence pion PDF through a deep neural network technique is presented in \sec{DNNITD}. In \sec{hybrid}, we discuss the PDF from the hybrid-scheme renormalization and $x$-spacing matching. Finally, \sec{conclusion} contains our conclusions. Some technical aspects of the calculations including the NNLO matching and the DNN technique are discussed in the Appendix.

\section{Lattice setup}\label{sec:latset}

\begin{table}
\centering
\begin{tabular}{|c|c|c|c|c|c|c|}
\hline
\hline
Ensembles &  $m_\pi$  & $t_s$ &  $z/a$ & $n_z$ & \#cfgs & (\#ex,\#sl) \cr
$a,L_t\times L_s^3$ & (GeV) & & & & &\cr
\hline
$a=0.076$ fm, & 0.14 &6,8,& [0,32] & 0,1,2,3  & 350 & $(5, 100)$\cr
    \cline{5-7}
$64\times 64^3$ &    &10& &  4,5,6,7  & 350 & $(10,200)$\cr  
\hline
\hline
\end{tabular}
\caption{Details of the setup of the $a=0.076$ fm lattice used in this paper. The number of gauge configuration (\#cfgs) and the number of exact and sloppy inversion samples (\#ex,\#sl) are shown. The momentum are given in lattice units and can be computed by $P_z=2\pi n_z/(L_s a)$. 
}
\label{tb:setup1}\end{table}

In this paper, the new addition is the data set at $a=0.076$ fm with a physical pion mass, and we describe the details for this data set below.
To extract the bare matrix elements of the pion, we computed the pion two-point and three-point functions using the Wilson-Clover action 
with HYP smeared gauge fields for the valence quarks on a 2+1 flavor Highly Improved Staggered Quark (HISQ)~\cite{Follana:2006rc} action generated by the HotQCD collaboration~\cite{Bazavov:2019www} with a pion mass of 140 MeV and lattice spacing a = 0.076 fm. The lattice extent $L_t\times L_s^3$ is $64\times 64^3$. We used the tree-level tadpole improved result for the coefficients of the clover term, $c_{sw}=u_0^{-3/4}=1.0372$, and the quark mass has been tuned so that the valence pion mass is 140 MeV as shown in \tb{setup1}. 
Here $u_0$ denotes the expectation value of the plauquette on HYP
smeared gauge configurations.
The gauge link entering the bilocal quark-bilinear operator has been 1-HYP smeared for an improved signal. This setup has recently 
been used in the calculations of the pion form factor~\cite{Gao:2021xsm} 
and pion distribution amplitude~\cite{Gao:2022vyh}.

The calculations were performed on GPUs, with the QUDA multigrid algorithm~\cite{Brannick:2007ue, Clark:2009wm, Babich:2011np, Clark:2016rdz} used for the Wilson-Dirac operator inversions to get the quark propagators. We used the All Mode Averaging (AMA)~\cite{Shintani:2014vja} technique to increase the statistics with a stopping criterion of $10^{-10}$ and $10^{-4}$ for the exact (ex) and sloppy (sl) inversions respectively. One of the crucial ingredients to access the parton distribution functions of the pion is the large momentum.  To obtain an acceptable signal for
pions at large momentum it is necessary to use
boosted sources \cite{Bali:2016lva}.
With the setup above, we are able to achieve momentum as large as 1.78 GeV for the physical pion mass with a reasonable signal quality. To control the lattice spacing and pion mass dependence, we combine the analysis with two other ensembles with a 300 MeV pion mass, lattice spacing a = 0.04, 0.06 fm and largest momentum up to 2.42 GeV, which has been discussed in detail and analyzed in \cite{Izubuchi:2019lyk,Gao:2020ito}.

\subsection{Pion two-point functions}

To constuct the boosted pion state, we compute the pion two point function,
\begin{align}\label{eq:c2pt}
\begin{aligned}
&C^{ss'}_{\rm 2pt}(t_s;P_z)=\left\langle \pi_s(\mathbf{x_0},t_s) \pi_{s'}^\dagger(\mathbf{P},0) \right\rangle ,
\end{aligned}
\end{align}
using the standard pion operator projected to spatial momentum $\mathbf{P}$,
\begin{align}\label{eq:pisource}
\begin{aligned}
&\pi_s(\mathbf{x},t_s)=\overline{d}_s(\mathbf{x},t_s)\gamma_5 u_s(\mathbf{x},t_s),\\ 
& \pi_s(\mathbf{P},t_s)= \sum_{\mathbf{x}} \pi_s(\mathbf{x},t_s)  e^{-i \mathbf{P}\cdot\mathbf{x}}.
\end{aligned}
\end{align}
The subscript $s$ in $\pi_s$ indicates the choice of quark smearing. In this work, 
we only work with the choice $\mathbf{P}=(0,0,P_z)$.
Since we use periodic boundary conditions, the hadron momentum is given by
\begin{align}
    P_z = \frac{2\pi n_z}{L_sa}
\end{align}
with $n_z$ listed in \tb{setup1}. The operator will create infinitely-many hadron states with the same quantum numbers as the pion, which are not only the pion ground states but also the excited states. We used momentum-smeared Gaussian-profiled sources in the Coulomb gauge~\cite{Izubuchi:2019lyk} $\pi_s(\mathbf{P},t)$ ($s=S$) to increase the overlap with the pion ground state, and to improve the signal for large momentum. We constructed smeared-smeared correlators ($s=s'=S$ labeled as SS) and smeared-point correlators ($s=S,s'=P$ labeled as SP) to decompose the energy levels of the pion spectrum. We tuned the radius of the Gaussian profile to be 0.59 fm for the $a = 0.076$ fm ensemble. For the boosted smearing, we used the quark momentum $n_z^k$ = 2 and 5 for hadron momentum $n_z\in[0, 3]$ and $[4, 7]$ respectively. Since the different pion momentum $n_z$ with the same quark momentum share the forward propagator, this enabled us to save some computation time. The details of the other two ensembles can be found in \refcite{Gao:2020ito}.

\subsection{Pion three-point functions}

We extracted the bare matrix elements from the three-point functions
\begin{align}\label{eq:c3pt}
C_{\rm 3pt}(z,\tau,t_s)=\left\langle \pi_S(\mathbf{x_0},t_s) O_\Gamma(z,\tau) \pi_S^\dagger(\mathbf{P},0)\right\rangle,
\end{align}
only using the smeared source and sink pion operator separated by a Euclidean time $t_s$. The iso-vector operator $O_\Gamma(z,\tau)$ inserted at time slice $\tau$ is defined as
\begin{align}
\begin{aligned}
O_\Gamma(z,\tau)&=&\sum_{\mathbf{x}}\bigg{[}\overline{u}(x+{\cal L})\Gamma W_z(x+{\cal L},x) u(x)-\cr &&\overline{d}(x+{\cal L})\Gamma W_z(x+{\cal L},x) d(x) \bigg{]},
\end{aligned}
\end{align}
with $x=(\mathbf{x}, \tau)$. The quark-antiquark pairs are separated along the momentum direction ${\cal L}=(0,0,0,z)$, and the Wilson line $W_z(x+{\cal L},x)$ makes sure such measurements are gauge invariant. In this work, we consider the choice  $\Gamma=\gamma_0$, since it reduces
reduces the higher-twist contribution~\cite{Radyushkin:2017cyf} and is free of mixing due to $\mathcal{O}(a^0)$ chiral symmetry breaking~\cite{Constantinou:2017sej,Chen:2017mie}. The gauge-links that enter the Wilson-line were 1-HYP smeared.

\section{Analysis of pion two-point functions}\label{sec:c2pt}

To extract the pion ground state matrix elements, one needs to know the energy levels created by the pion operator, which can be determined by analyzing the two-point functions. 
Below, we discuss the analysis of the two-point function for the $a = 0.076$ fm, physical quark mass ensemble.
The details for the spectral analysis of the other two ensembles with a 300 MeV pion mass are given in \refcite{Gao:2020ito}.  The pion two-point functions constructed from \Eq{c2pt} have the spectral decomposition,
\begin{align}\label{eq:spectr}
    C^{ss'}_{\rm 2pt}(t_s)=\sum_{n=0} A^s_n A_n^{s'*} (e^{-E_n t_s}+e^{-E_n (aL_t-t_s)}),
\end{align}
with $E_{n+1}\textgreater E_{n}$ being the energy level and, $A^s_n=\langle \Omega|\pi_s|n\rangle$, is the pion overlap amplitude. Here $|\Omega\rangle$ denotes the vacuum state.

\begin{figure}
	\includegraphics[width=0.35\textwidth]{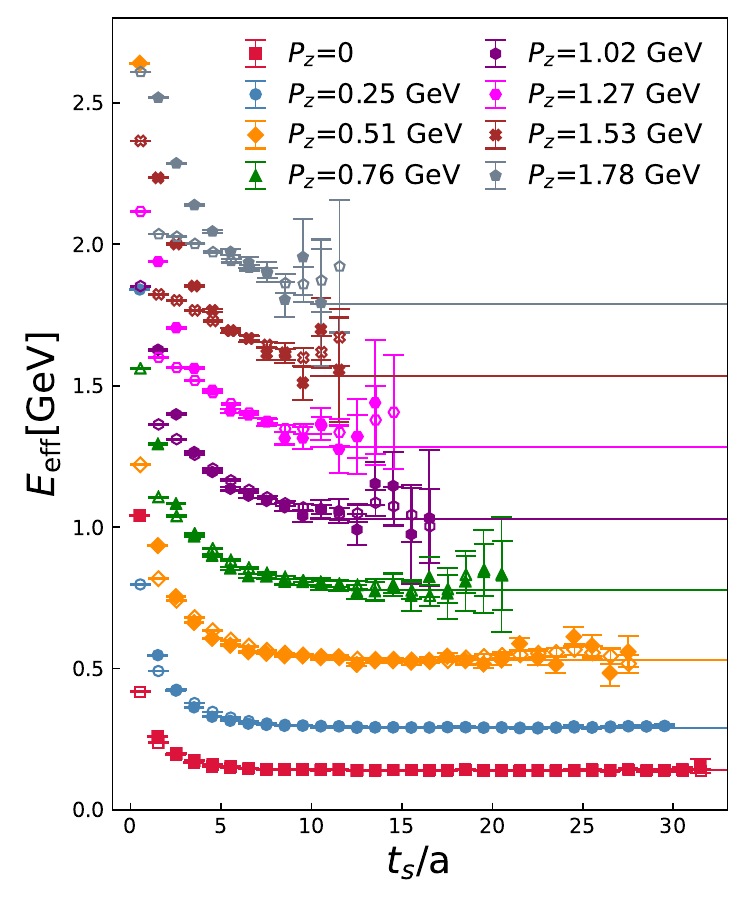}
	\caption{Pion effective masses for the $m_\pi$ = 140 MeV ensemble are shown. The filled and open symbols correspond to the SS and SP correlators respectively. The lines are calculated from dispersion relation $E(P_z)=\sqrt{P_z^2+m_\pi^2}$ with $m_\pi$ = 140 MeV.\label{fig:effmassHISQ}}
\end{figure}

\begin{figure}
	\includegraphics[width=0.23\textwidth]{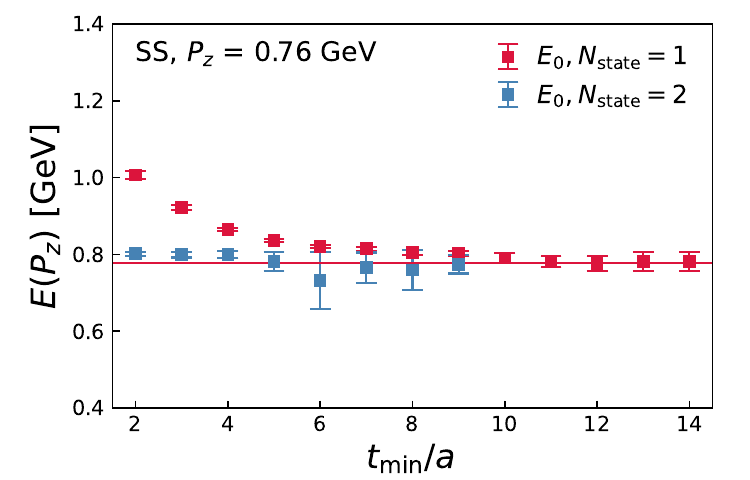}
	\includegraphics[width=0.23\textwidth]{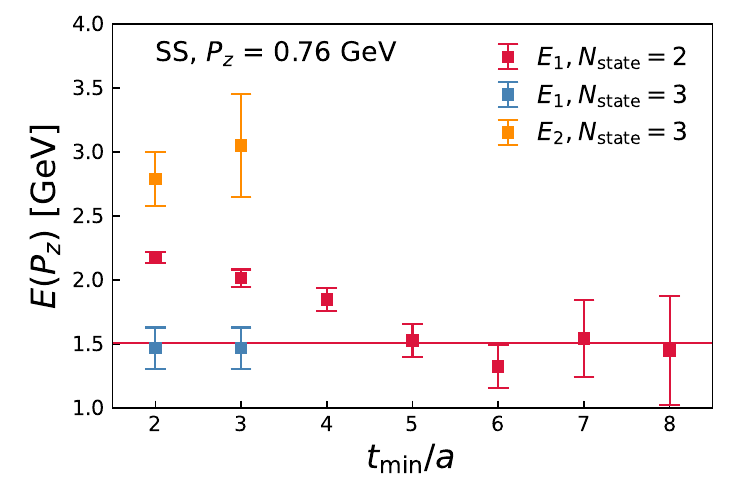}
	\includegraphics[width=0.23\textwidth]{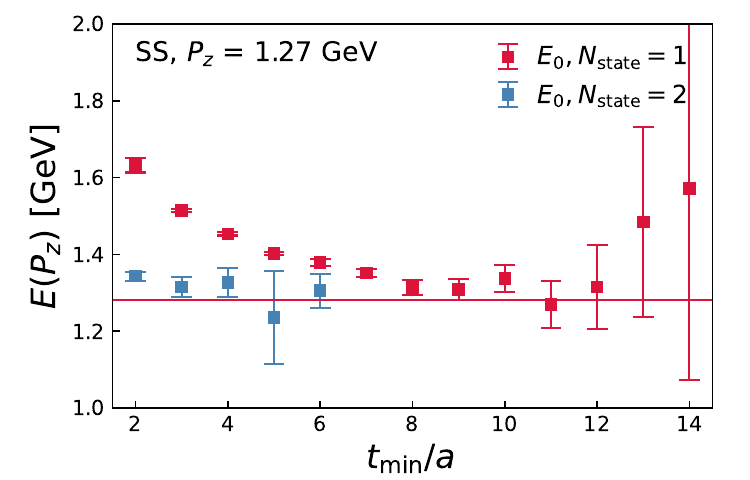}
	\includegraphics[width=0.23\textwidth]{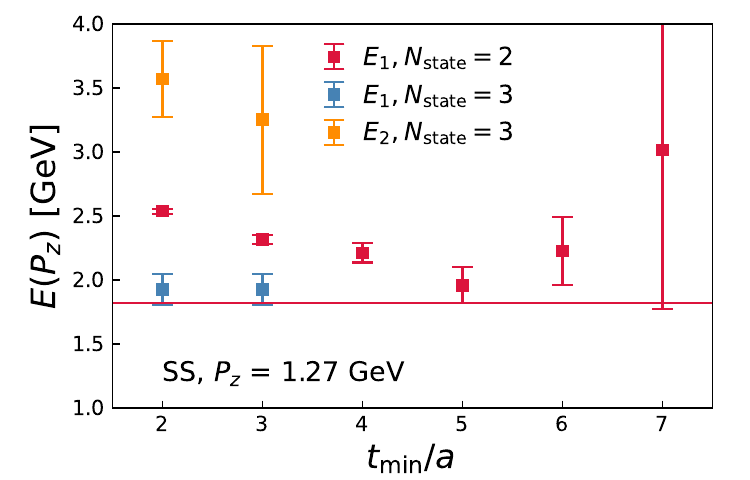}
	\caption{$N$-state fit results for $n_z=3,5$ on the $m_\pi$ = 140 MeV ensemble. The left panels show $E_0$ from one- and two-state fits to the SS correlators, while the right panels show $E_1$ from constrained two- and three-state fits to the SS correlators. The lines in left panels are computed from dispersion relation $E_0(P_z)=\sqrt{P_z^2+m_\pi^2}$ with $m_\pi$ = 0.14 GeV, while the ones in the right panels are with $m_\pi'$ = 1.3 GeV.\label{fig:c2ptfitHISQ}
 }
\end{figure}

It is expected that the boosted Gaussian smeared sources should have a good overlap with the pion ground state. In \fig{effmassHISQ}, we show the effective mass of the SS (filled symbols) and SP (open symbols) correlators from the $a = 0.076$ fm ensemble. At very small $t_s$, the effective masses corresponding to the SS correlators are larger than the ones
corresponding to SP correlator. This is likely due to the effect that some higher excited states contribute with a negative weight to the SP correlator.
However, the ordering of the effective masses changes at larger $t_s$ the ordering changes and
the effective mass of SS correlators reaches a plateau around $t_{s} \lesssim 10a$. 
For the case of $P_z=0$, we have the pion mass at the physical point around 140 MeV. The lines in the figure are calculated from the dispersion relation $E_0(P_z)=\sqrt{P_z^2+m_\pi^2}$ with $m_\pi$ = 140 MeV, which show nice agreement with the plateaus. The different behavior of the SS and SP correlators can help the extraction of the excited energy states.

We truncated \Eq{spectr} up to $n=N-1$ and performed an $N$-state fit to the two-point functions with time separations in the range $[t_{\textup{min}},32a]$. 
We show the fit results for $n_z=3,5$
in \fig{c2ptfitHISQ} as examples. In the left panels, we show the lowest energy $E_0$ 
of the SS correlators extracted from one-state (red points) and two-state (blue points) fits respectively. 
The values of $E_0$ from the one 
state fits show similar behavior to the effective mass (see \fig{effmassHISQ}) 
as a function of $t_{\textup{min}}$. They reach plateaus around $t_{s} \lesssim 10a$, and agree with the red lines calculated from the dispersion relation. 
From the two-state fits we find that the values of $E_0$ reach plateaus at smaller $t_s$, namely  $t_{\textup{min}}\sim3a$. 
Therefore, from one-state and two-state models of the pion two-point function we can extract the pion ground state 
using $t_{s} \ge 10a$ and $t_{s} \ge 3a$, respectively. 
To determine the first excited state $E_1$, we fixed $E_0$ to be the best estimate from the one-state fit and performed a constrained two-state fit. 
The values of $E_1$ from the SS correlators are shown as red points in the right panels of \fig{c2ptfitHISQ}. These values are slowly decreasing  with increasing $t_{\textup{min}}$ 
and roughly reach a plateau around $t_{\textup{min}}\sim 5a$ within the statistical errors. 
The plateaus are consistent with the red lines calculated from the dispersion relation $E_1(P_z)=\sqrt{P_z^2+m'^2_\pi}$ with $m'_\pi\approx$ 1.3 GeV, 
which may suggest a single paricle state~\cite{Gao:2021hvs}. To decompose the higher energy states through the three-state fit of the SS correlators, 
in addition to fixing the $E_0$, we also added a prior on $E_1$ from the best estimates from the SP correlators with corresponding errors~\cite{Gao:2020ito}. 
The $E_1$ (blue points) and $E_2$ (orange points) extracted from the constrained three-state fit of the SS correlators are shown in the right panels of \fig{c2ptfitHISQ}. 
The three-state fit only works for $t_{\textup{min}}\le 3a$ because of the limited statistics, and the $E_2$ values do not show $t_{\textup{min}}$ dependence within errors. 
But, instead of a single-particle state, the $E_2$ is more likely to be the combination of the tower of energies beyond $E_1$ which however cannot be decomposed with the limited statistics. 
From this, we conclude that the spectrum present in the smeared-smeared (SS) correlators can be described by a three-state model 
when $t_{\textup{min}}\gtrsim 2a$ and by a two-state model for $t_{\textup{min}} \gtrsim 5a$. 
These facts will contribute to the analysis of the three-point functions in the next section.
\section{The ground state bare matrix elements from three-point functions}\label{sec:c3pt}

\begin{figure*}
	\includegraphics[width=0.32\textwidth]{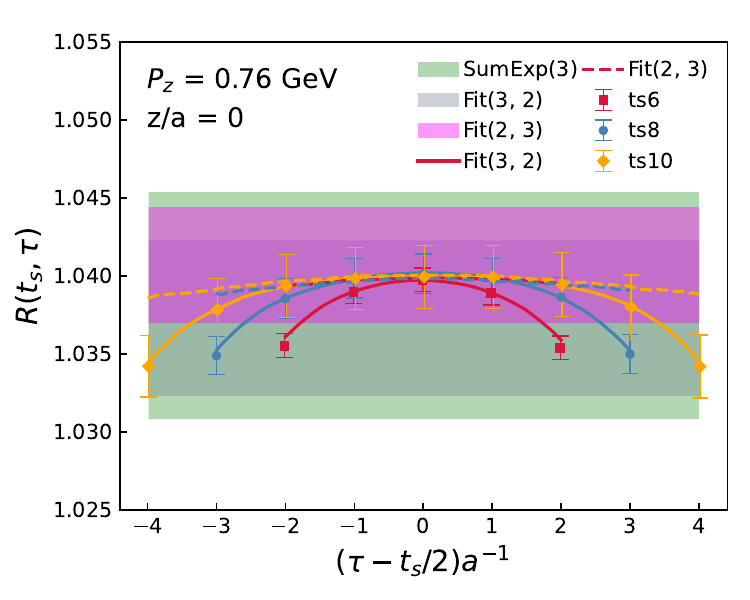}
	\includegraphics[width=0.32\textwidth]{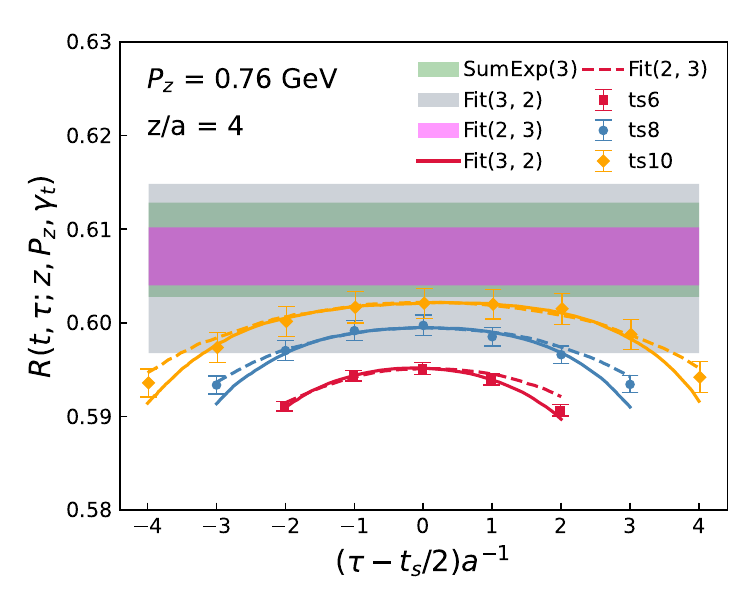}
	\includegraphics[width=0.32\textwidth]{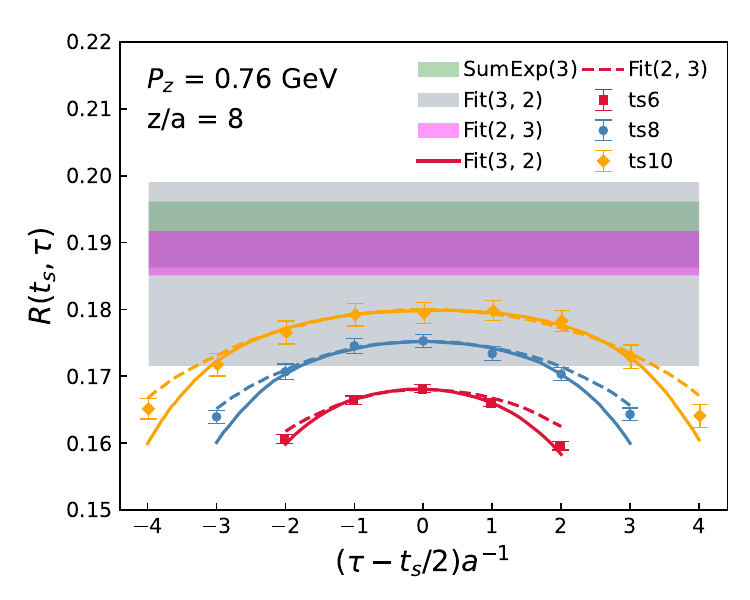}
	\includegraphics[width=0.32\textwidth]{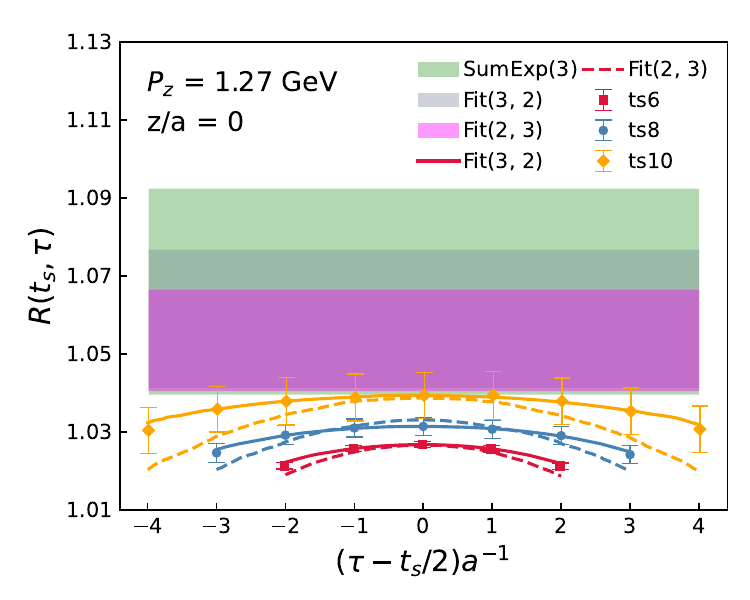}
	\includegraphics[width=0.32\textwidth]{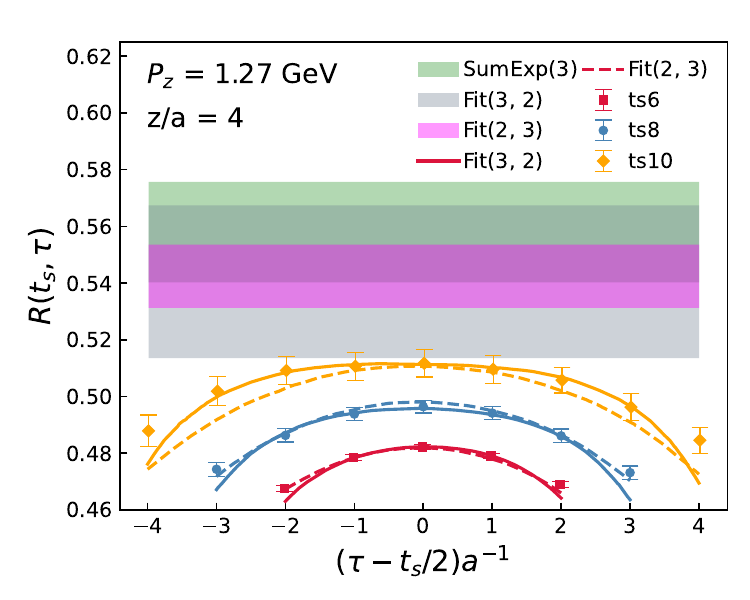}
	\includegraphics[width=0.32\textwidth]{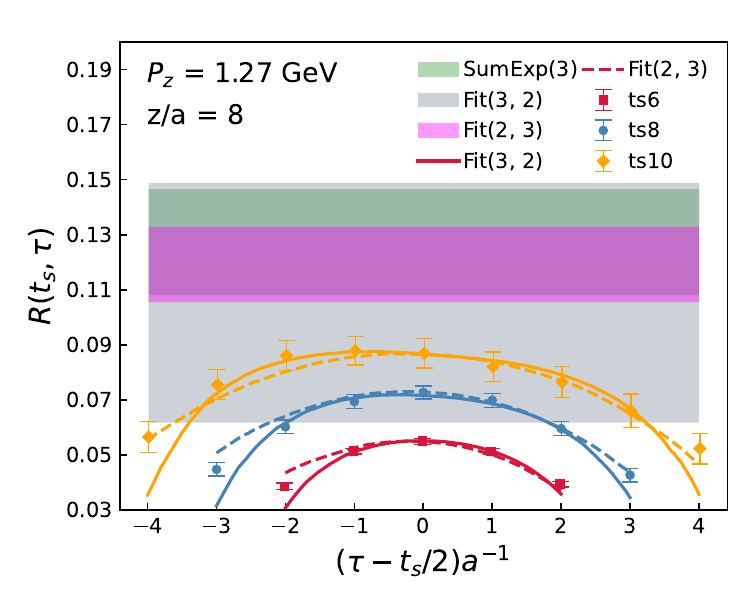}
	\caption{Ratios $R(t_s,\tau)$ for $n_z$ = 3 (upper panels), 5 (lower panels) of the $m_\pi$ = 140 MeV ensemble are shown. The continuous and dashed curves are the central values from Fit(3,2) and Fit(2,3) respectively. The horizontal bands differentiated by their colors are the fit results of SumExp(3), Fit(3,2) and Fit(2,3).\label{fig:ratioHISQ}}
\end{figure*}

\begin{figure}
	\includegraphics[width=0.23\textwidth]{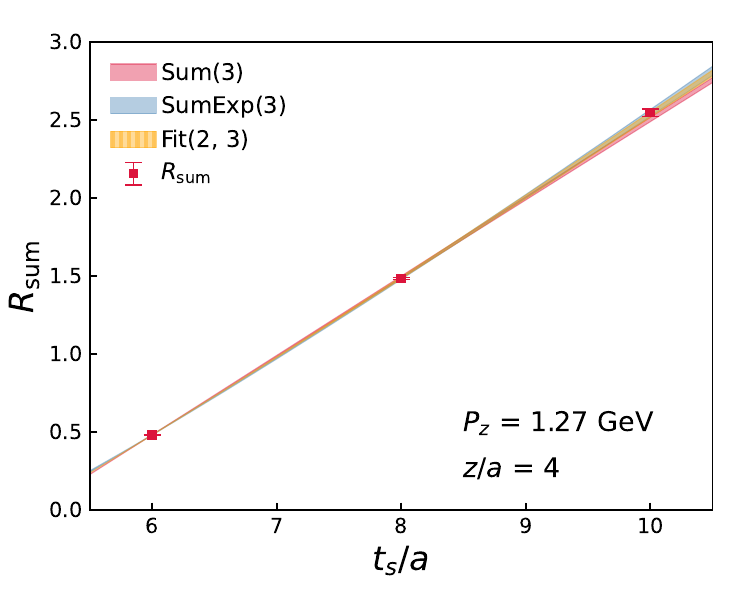}
	\includegraphics[width=0.23\textwidth]{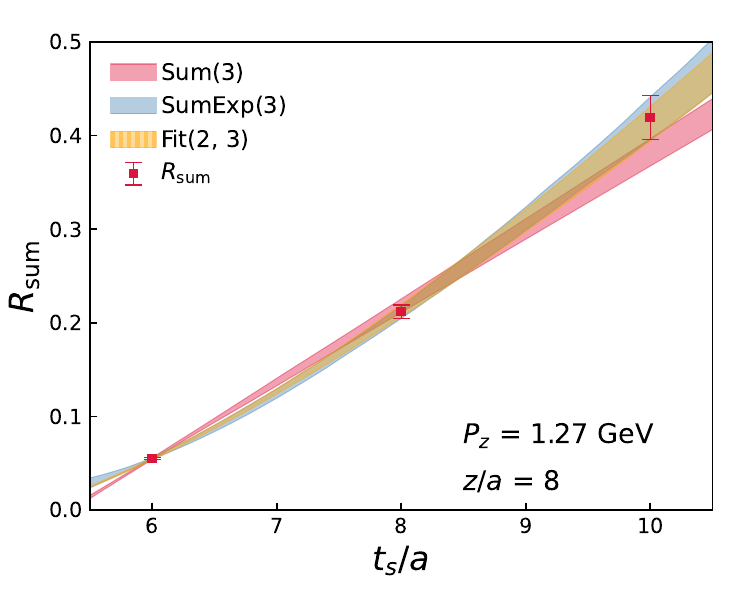}
	\caption{$R_{\textup{sum}}(t_s)$ for $n_z$ = 5 with $z/a$ = 4 and 8 is shown. The bands are reconstructed from fit strategies Sum(3), SumExp(3), and Fit(2,3).\label{fig:ratiosum}}
\end{figure}

The three-point function defined in \Eq{c3pt} has the spectral decomposition,
 \begin{align}\label{eq:c3ptsp}
 \begin{aligned}
&\left\langle \pi_S(\mathbf{x}_0,t_s) O_\Gamma(z,\tau) \pi_S^\dagger(\mathbf{P},0)\right\rangle \\
&= \sum_{m,n}\langle \Omega|\pi_S|m\rangle\langle m|O|n\rangle\langle n|\pi_S^\dagger|\Omega_\Gamma\rangle e^{-\tau E_{n}}e^{-(t_s-\tau)E_m},
 \end{aligned}
\end{align}
where the overlap amplitudes $A^s_n=\langle \Omega|\pi_S|n\rangle$ and the energy levels $E_n$ are, by definition, the same as for the two-point functions. The quantity $h^B(z,P_z)=\langle 0|O_\Gamma|0\rangle$ is the bare matrix element of the pion ground state. 
We only computed the three-point function with both the pion source and sink smeared.
To take advantage of the high correlation between the three-point and two-point functions, we construct the ratio
\begin{align}\label{eq:ratio}
R(t_s,\tau)=\frac{C_{\textup{3pt}}(t_s,\tau)}{C^{SS}_{\textup{2pt}}(t_s)},
\end{align}
which in the limit $t_s\to\infty$ approaches $R(t_s,\tau)=\langle 0|O_\Gamma|0\rangle$.
To take care of the possible wrap around effect, we found it 
convenient to replace
$C^{SS}_{\textup{2pt}}(t_s)$ with $C^{SS}_{\textup{2pt}}(t_s)-|A_0|^2e^{-E_0t_s}$ using the value of $E_0(P_z)$ from the best estimate of \sec{c2pt}. 
We applied the following two methods to extract the bare matrix elements $h^B(z,P_z)$:
\begin{enumerate}
  \item As shown in \Eq{spectr} and \Eq{c3ptsp}, the spectral decomposition is known for both the three-point and two-point functions. Therefore, we can apply the $N$-state fits taking the values of $|A_n|^2$ and $E_n$ from the fits of \sec{c2pt} to extract the bare matrix elements $h^B(z,P_z)$ of the pion ground state. We will refer to this method by Fit($N$, $n_{\textup{sk}}$), with $N$ denoting the $N$-state fit and $n_{\textup{sk}}$ being the number of $\tau$ (time insertions) we skipped on two sides of each time separation.
  \item The sum of the ratios
  \begin{align}
      R_{\textup{sum}}(t_s)=\sum_{\tau=n_{\textup{sk}}a}^{t_s-n_{\textup{sk}}a}R(t_s,\tau),
  \end{align}
  has less excited-state contributions, and its dependence on $t_s$ can be approximated by a linear behavior~\cite{MAIANI1987420},
  \begin{align}
  \begin{aligned}
  &R_{\textup{sum}}(t_s)=(t_s-2n_{\textup{sk}}a)h^B(z,P_z)\\
  &+B_0+\mathcal{O}(e^{-(E_1-E_0)t_s}).
  \end{aligned}
  \end{align}
  However, when $t_s$ is too small, the excited-state contributions cannot be entirely neglected and one should include the leading-order correction~\cite{Gao:2020ito} of the form,
  \begin{align}
    \begin{aligned}
  &R_{\textup{sum}}(t_s)=(t_s-2n_{\textup{sk}}a)h^B(z,P_z)\\
  &+B_0+B_1e^{-(E_1-E_0)t_s}+\mathcal{O}(e^{-(E_2-E_0)t_s}).
   \end{aligned}
  \end{align}
  We will refer to the above two summation methods by Sum(n$_{\textup{sk}}$) and SumExp($n_{\textup{sk}}$) with $n_{\textup{sk}}$ being the number of $\tau$ we skipped.
\end{enumerate}

\begin{figure}
\centering
	\includegraphics[width=0.4\textwidth]{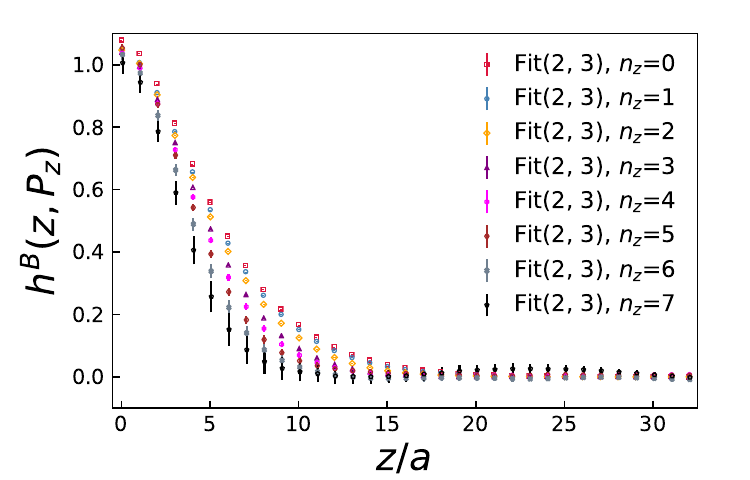}
	\caption{The bare matrix elements $h^B(z,P_z)$ of the $m_\pi$ = 140 MeV ensemble extracted from the Fit(2,3) method are shown for $P_z$ from 0 to 1.78 GeV.\label{fig:bmHISQ}}
\end{figure}

As discussed in \sec{c2pt}, a three-state spectral model can describe two-point functions with $t_s \geq 2a$, thus Fit(3,2) is justified to do the extrapolation of $R(t_s,\tau)$. In addition, we observed that $E_0$ from the two-state fit reaches the plateau around $t_{\textup{min}}\sim$ $3a$ and $4a$, where $E_1$ is consistent within errors as an effective description of the tower of energies larger than $E_0$. Therefore, we also tried the two-state fit as Fit(2,3) to reduce the number of fit parameters. In \fig{ratioHISQ}, we show the ratios $R(t_s,\tau)$ for $n_z=3,5$ with $z/a=0,4,8$ as examples, where the central values of the fit for Fit(3,2) and Fit(2,3) are shown as solid and dashed curves, both describing the $R(t_s,\tau)$ data well within the errors. We tried both summation methods Sum(3) and SumExp(3), as shown in \fig{ratiosum}. We show the fit result of the two summation methods and also reconstruct the corresponding bands from the two-state fit Fit(2,3) for comparison. As one can see, SumExp(3) can better describe the data as compared to Sum(3) and it is consistent with Fit(2,3), suggesting that the exited-state contamination cannot be totally neglected in the summation methods. Therefore, we only used the corrected summation fit SumExp($n_{\textup{sk}}$) and abandoned Sum($n_{\textup{sk}}$) in our analysis. Finally, we compare the extracted bare matrix elements from the $N$-state fits and SumExp(3), shown as horizontal bands in \fig{ratioHISQ}, where consistent results can be observed indicating the robustness of our analysis. The results from Fit(3,2) are usually noisier due to having 9 fit parameters, whereas Fit(2,3) only has 4 fit parameters. Since the summation methods are the approximated form of Fit(2,3), we will take the results of Fit(2,3) for our subsequent analysis. To summarize, the bare matrix elements for all momenta using the Fit(2,3) method are shown in \fig{bmHISQ}. 
The local bare matrix element $h^B(z=0,P_z)$, which is the vector current renormalization factor $Z_V$, has been discussed in \refcite{Gao:2021xsm}.
\section{Ratio-scheme renormalization and leading-twist expansion}\label{sec:matching}

In \sec{c3pt} we extracted the bare matrix elements of the pion $h^B(z,P_z,a)$, which then need to be renormalized. It is known that the operator $O_\Gamma(z)$ (c.f. \Eq{operator}) is multiplicatively renormalizable~\cite{Ji:2017oey,Green:2017xeu,Ishikawa:2017faj},
\begin{align}
    h^B(z,P_z,a)=e^{\delta m|z|}Z(a)h^R(z,P_z,\mu),
\end{align}
with $Z(a)$ coming from the
fields and vertices renormalization, and $e^{\delta m|z|}$ coming from the self-energy divergence of the Wilson line.  Nonperturbative renormalization (NPR) schemes such as RI-MOM~\cite{Constantinou:2017sej,Alexandrou:2017huk,Chen:2017mzz,Stewart:2017tvs, Fan:2020nzz, Lin:2021brq, Lin:2020fsj, Lin:2020rxa} and the Hybrid scheme~\cite{Ji:2020brr, Gao:2021dbh, LatticePartonCollaborationLPC:2021xdx, Hua:2020gnw} are one possible 
way to remove the UV divergences. 
Alternatively, the renormalization factors, including the Wilson line renormalization, cancel out in the ratios of hadron matrix elements evaluated
at different momenta, $P_z$. That is, we 
construct the renormalization group invariant (RGI) ratios,
\begin{align}\label{eq:ratioR}
\begin{split}
   \mathcal{M}(z,P_z,P_z^0)
=\frac{h^B(z,P_z,a)}{h^B(z,P_z^0,a)}
=\frac{h^R(z,P_z,\mu)}{h^R(z,P_z^0,\mu)},
\end{split}
\end{align}
where the matrix elements in the rightmost term above are renormalized in the $\overline{\rm MS}$ scheme due to the ratio being RGI.
For $P_z^0=0$, the ratio is usually called the reduced Ioffe-time 
distribution (rITD)~\cite{ Radyushkin:2017cyf, Orginos:2017kos}, and the generalization of such a ratio via the 
use of non-zero $P_z^0$ was advocated in Ref~\cite{Gao:2020ito}.
Due to Lorentz invariance, $ \mathcal{M}(z,P_z,P_z^0)$ can be rewritten as $\mathcal{M}(z^2, zP_z,z P_z^0)$, with the term 
$\lambda=z P_z$ usually referred to as the Ioffe-time in the 
literature~\cite{ Radyushkin:2017cyf}.
For small values of $z$, one can use the leading-twist (twist-2) expansion to relate $\mathcal{M}(z,P_z,P_z^0)$ to the PDF, $q(x,\mu)$, as
~\cite{Radyushkin:2017cyf, Orginos:2017kos, Izubuchi:2018srq, BALITSKY1989541}
\begin{align}\label{eq:ratSDF}
\begin{split}
&\mathcal{M}(z,P_z,P_z^0)=\\
&\displaystyle \frac{\sum_{n=0} c_n(\mu^2z^2)\frac{(-izP_z)^n}{n!}\langle x^n \rangle(\mu) +
{\cal O}(\Lambda_{\rm QCD}^2 z^2)}{\sum_{n=0} c_n(\mu^2z^2)\frac{(-izP_z^0)^n}{n!} \langle x^n \rangle(\mu) + {\cal O}(\Lambda_{\rm QCD}^2 z^2) },
\end{split}
\end{align}
where $c_n(\mu^2z^2)=C_n(\mu^2z^2)/C_0(\mu^2z^2)$, and $C_n(\mu^2z^2)$ are the Wilson coefficients calculated from perturbation theory. The Mellin moments are defined as
\begin{align}
    \langle x^n\rangle(\mu) = \int_{-1}^1dx\, x^nq(x,\mu) ,
\end{align}
with $q(x,\mu)$ being the light-cone PDF. The Wilson coefficients have been calculated at NLO \cite{Izubuchi:2018srq,Radyushkin:2017lvu} as well as at NNLO \cite{Chen:2020ody,Li:2020xml}. It can be seen that the matrix elements in both numerator and denominator suffer from the higher-twist effects ${\cal O}(\Lambda_{\rm QCD}^2 z^2)$, which are less important when $P_z^0, P_z>\Lambda_{\rm QCD}$. The ratio-scheme renormalization defined above has the potential to extend the range of $z$ by reducing the higher-twist effects ${\cal O}(\Lambda_{\rm QCD}^2 z^2)$ through the possible cancellation between the numerator and denominator. It was observed in \refcite{Gao:2020ito} that the difference  in the resulting PDF when using different $P_z^0$ in the ratio scheme is marginal within the errors. Since the lattice results of $h^B(z,P_z,a)$ for different values of $z$ are strongly correlated in practical calculations, it is convenient to replace the ratio defined by Eq. (\ref{eq:ratioR}) with the following one
\begin{align}\label{eq:ratmod}
   \mathcal{M}(z,P_z,P_z^0)
=\frac{h^B(z,P_z,a) h^B(0,P_z^0,a)}{h^B(z,P_z^0,a) h^B(0,P_z,a)}\,.
\end{align}
This ratio is equivalent to the one in Eq. (\ref{eq:ratioR}) in the continuum limit, but it achieves two things. First, it imposes the 
condition that the value of the $z=0$ matrix element is momentum independent and therefore removes certain lattice corrections. Second, such a modified ratio has smaller statistical errors
because of the correlations between the $z=0$ and $z\ne 0$ matrix elements.
Therefore, we will use Eq. (\ref{eq:ratmod}) to estimate $\mathcal{M}(z,P_z,P_z^0)$ from our lattice
calculations.
\begin{figure}
\centering
	\includegraphics[width=0.4\textwidth]{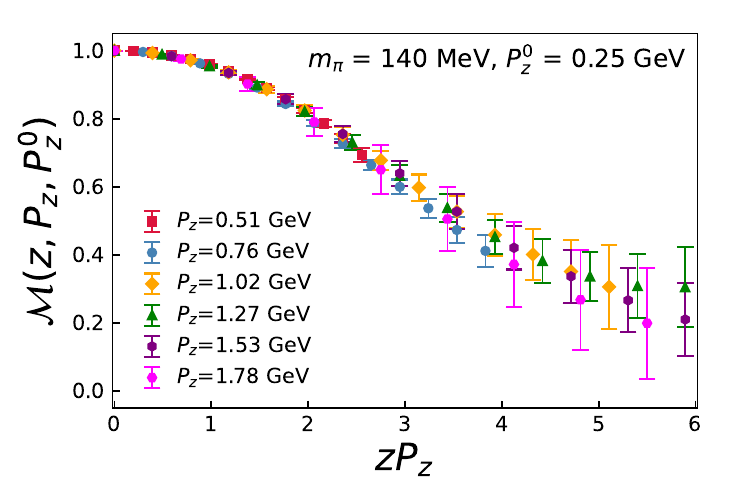}
	\caption{The ratio scheme renormalized matrix elements $\mathcal{M}(z,P_z,P_z^0)$ with $P_z^0=0.25$ GeV are shown for the $m_\pi$ = 140 MeV ensemble.\label{fig:ritdpznm1}}
\end{figure}

The matrix element $h^B(z,P_z)$ could be affected by the finite temporal extent of the lattice, $L_t$~\cite{Briceno:2018lfj,Lin:2019ocg,Liu:2020krc}. These wrap-around effects that are proportional to $e^{-m_\pi L_t}$ could be as large as 3\% for $P_z=0$ and $m_{\pi} L_t \le 3.5$. Such wrap-around effect was discussed and estimated in \refcite{Gao:2020ito, Gao:2021xsm}. It was found to be important for zero momentum case but negligible for the cases of $P_z > 0$. Thus, the use of $P_z^0\ne 0$ offers yet another practical advantage. For this reason, for the $m_\pi=$ 140 MeV ensemble, we consider $P_z^0=2 \pi/L_s \simeq 0.25$ GeV and omit the $P_z=0$ lattice data from the analysis in what follows. Finally, in Fig. \ref{fig:ritdpznm1}, we show our results for $\mathcal{M}(z,P_z,P_z^0)$ with $P_z^0=0.25$ GeV as a function of $\lambda =z P_z$, and different values of $P_z$.

\section{Model independent determination of even moments of the pion valence quark PDF}\label{sec:moments}

\subsection{Leading-twist (twist-2) OPE and the fitting method}

\begin{figure}
	\includegraphics[width=0.4\textwidth]{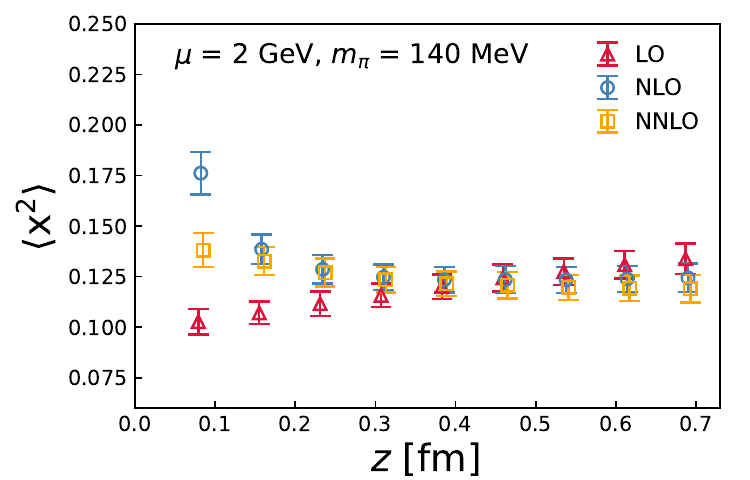}
	\caption{$\langle x^2 \rangle$ extracted from the ratio-scheme data $\mathcal{M}(z,P_z,P_z^0)$ at each $z$ using LO, NLO, and NNLO kernels is shown for the $m_\pi$ = 140 MeV ensemble.\label{fig:momsingle076}}
\end{figure}

The leading-twist expansion formula given by Eq. (\ref{eq:ratSDF}) tells us that $\mathcal{M}(z,P_z,P_z^0)$ is sensitive to the moments of the PDF, and these can be extracted by fitting the $z P_z$ and $z^2$ dependencies of the lattice data for the ratio. As discussed in \sec{latset}, we are interested in the iso-vector PDF $q^{u-d}(x)$ of the pion, which is defined on $x\in$ [-1, 1] and obeys $\int_{-1}^1dx\,q^{u-d}(x)=1$. Under isospin symmetry, the pion iso-vector PDF $q^{u-d}$ is symmetric with respect to $x$=0 and is equivalent to the pion valence PDF defined on $x\in$ [0, 1], namely $q^{u-d}(|x|)=q^{u-d}(-|x|)=q^v(|x|)$. Because of this, only the terms even in $n$ contribute to Eq. (\ref{eq:ratSDF}).
If we assume the positivity of the $\overline{\rm MS}$ pion valence PDF~\footnote{We are aware of Ref.~\cite{Collins:2021vke} which shows that positivity is not a necessary constraint for $\overline{\rm MS}$ PDFs (also see references therein). Nevertheless, we still use positivity as a constraint for pion valence quark PDF in our analysis.}, we have $\langle x^{n} \rangle>0$. Based on positivity, further constraints can also be imposed on the moments as~\cite{Gao:2020ito}
\begin{align}\label{eq:constr}
    \langle x^{n+2} \rangle &- \langle x^{n} \rangle < 0,\nonumber\\
    \langle x^{n+2} \rangle &+ \langle x^{n-2} \rangle - 2\langle x^{n} \rangle > 0.
\end{align}

In order to extract the moments of the PDF using Eq. (\ref{eq:ratSDF}) it is necessary to ensure that the leading-twist expansion is a good approximation for the data under consideration. This implies that $z$ cannot be too large, which given the fact that $P_z$ is less than 3 GeV in present day lattice calculations
also means that the range in $\lambda$ is limited. As we see from Fig. \ref{fig:ritdpznm1}, our lattice data cover the range up to $\lambda \simeq 6$ with $z$ up to around 0.608 fm. In this range, the sums in Eq. (\ref{eq:ratSDF}) can be truncated to a few terms. As discussed in Appendix \ref{app:momsTruncate}, for realistic pion PDF
in this range of $\lambda$ the sums 
can be truncated at $n_{\rm max}=8$. Here we note that there are also higher-twist terms entering the leading-twist formula that are proportional
to $(m_\pi^2z^2)^{n}$, known as target mass corrections. 
The target mass corrections can be taken care of by the following replacement~\cite{Chen:2016utp,Radyushkin:2017ffo}
\begin{align}
\langle x^n \rangle \rightarrow  \langle x^n \rangle \sum_{k=0}^{n/2}\frac{(n-k)!}{k!(n-2k)!}\Big(\frac{m^2_\pi}{4P_z^2}\Big)^k.
\end{align}
These corrections are small in the case of the pion. We have to ensure that the higher-twist corrections beyond the target mass corrections are also small. Furthermore, we need to ensure that the perturbative expression for $c_n$ is also reliable.
To do this we fit the $\lambda$ dependence
of $\mathcal{M}(z,P_z,P_z^0)$ at fixed $z$ and check to what extent the obtained moments are independent of $z$.
We perform  fits at each $z$ (6 data points with different $P_z$) by truncating the moments up to $n^{\rm max}$. To stabilize the fits, we impose constraints given by Eq. (\ref{eq:constr}). We tried $n^{\rm max} = 6$ and 8, where reasonable $\chi^2_{d.o.f} \lesssim$ 1 can always be found when $z\ge2a$ for $n^{\rm max}$ = 6. As for $n^{\rm max}$ = 8, we found $\langle x^{8} \rangle$ is always consistent with 0. We therefore fix $n^{\rm max} = 6$ for the following discussion in this section. 
We use LO, NLO and NNLO results for $c_n$ and fix the scale $\mu$ to 2 GeV. The coupling constant $\alpha_s(\mu)$ entering $c_n(\mu^2z^2)$ at NLO and NNLO is evolved from $\alpha_s(\mu=2\ {\rm GeV})=0.293$, which is obtained from $\Lambda_{\rm QCD}^{\overline{\rm MS}}=332$ MeV with the five-loop $\beta$-function and $n_f=3$, as has been calculated using the same lattice ensembles~\cite{Petreczky:2020tky}.

In \fig{momsingle076}, we show the second moment $\langle x^2 \rangle$ extracted from each fixed $z$ using LO, NLO as well as NNLO matching kernels for the $a = 0.076$ fm ensemble. Clear $z$ dependence can be observed at LO. Beyond LO, the perturbative kernels are supposed to compensate the $z$ dependence and produce the $z$-independent plateaus of $\langle x^2 \rangle(\mu)$. 
We see from the figure that for $z>0.3$ fm the NLO and NNLO results show no $z$-dependence within errors, suggesting
that the leading-twist approximation is reliable in this $z$-range. On the other hand, for smaller $z$ we see
a clear dependence on $z$, and there is a clear difference between the NLO and NNLO result. This is counter intuitive,
as one expects the short distance factorization to work better for small $z$. To understand this, we performed
additional studies using the results for $\mathcal{M}(z,P_z,P_z^0)$ calculated on the $a=0.04$ fm and $a=0.06$ fm ensembles. We varied
the renormalization scale roughly by a factor of $\sqrt{2}$, i.e. we used $\mu=1.4$ GeV and $\mu=2.8$ GeV, and also performed
an analysis where the coupling constant was evolved from the scale $\mu=2$ GeV to a scale proportional to $1/z$, which corresponds
to resumming logs, $\ln(\mu^2 z^2)$. These additional calculations are discussed in detail in Appendix \ref{app:dataevo}. There it is also pointed out that for $z/a \le 2$ the extractions of the moments is affected by lattice discretization effects.
The conclusion of the analysis presented in Appendix \ref{app:dataevo} is that the resummation of logs is important for
$z<0.1$ fm and the resummed expressions for $c_n(\mu^2 z^2)$ work well for $z\le 0.3$ fm. 
For large $z$ the resummed results for $c_n$ are not appropriate as the scale in the running coupling constant becomes too low.
However, the moments obtained using the resummed result for $c_n$ and $z<0.3$ fm agree with the ones obtained with the fixed order result and $z>0.3$ fm. Therefore it seems the fixed-order NNLO leading-twist approximation can describe the ratio-scheme data at least up to 0.7 fm, though it was not expected from the beginning. But this allows us to determine higher moments of the pion PDF and constrain the pion PDF in general, as we discuss later.

\subsection{First few moments from a combined fit}

\begin{figure}
	\includegraphics[width=0.4\textwidth]{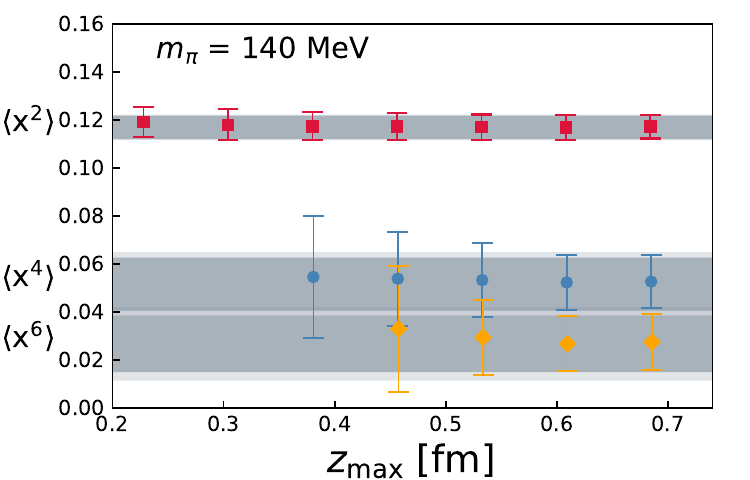}
	\caption{$\langle x^n \rangle$ with $n = 2,~4,~6$ extracted from the combined fit are shown for the $m_\pi=140$
	MeV lattice. The darker and lighter bands correspond to the statistical and systematic errors. The fits use the ratio-scheme data $\mathcal{M}(z,P_z,P_z^0)$ for $z\in[2a, z_{\rm{max}}]$ and $P_z > P_z^0=0.25$ GeV.\label{fig:momsCombHISQ}}
\end{figure}

\begin{table}
\centering
\begin{tabular}{c|c|c|c|c}
\hline
\hline
$a$ [fm]&$\langle x^2 \rangle$ & $\langle x^4 \rangle$& $\langle x^6 \rangle$ & $\chi^2_{d.o.f}$\cr
\hline
0.076 & 0.1169(49)(03) & 0.0516(110)(19) &0.0267(118)(28) &0.91 \cr
\hline
0.06 & 0.1159(35)(09) & 0.0385(61)(32) &0.0139(73)(41) & 1.4 \cr
\hline
0.04 & 0.1108(38)(05) & 0.0396(46)(18) &0.0146(43)(16) & 0.63\cr
\hline
\hline
$a^2\rightarrow$0 & 0.1104(73)(48) & 0.0388(46)(57) &0.0118(48)(48) &1.3 \cr
\hline
\hline
\end{tabular}
\caption{$\langle x^n \rangle$ with $n = 2,~4,~6$ extracted from the combined fit are shown for the three ensembles at $\mu$ = 2 GeV. The statistical errors are in the first brackets, while the systematic errors are in the second brackets which are estimated by varying  $z_{\rm max}\in[0.48, 0.72]$ fm for the fits.}
\label{tb:momsHISQ}\end{table}

To stabilize the fit and extract the higher moments, we perform a combined fit over a range of ratio scheme data in [$z_{\rm min}$, $z_{\rm max}$]. As has been mentioned, we will use the ratio scheme data $\mathcal{M}(z,P_z,P_z^0)$ constructed from $P_z~ \textgreater ~P_z^0=0.25$ GeV. To take care of possible cut-off effects, we apply the modified form~\cite{Gao:2020ito},
\begin{align}\label{eq:ratioTMClat}
\begin{aligned}
\mathcal{M}(z,P_z,P_z^0)=\frac{\sum_nc_n(\mu^2z^2)\frac{(-izP_z)^n}{n!}\langle x^n \rangle + r(aP_z)^2}{\sum_nc_n(\mu^2z^2)\frac{(-izP_z^0)^n}{n!} \langle x^n \rangle + r(aP_z^0)^2},
\end{aligned}
\end{align}

The fit results are shown in \fig{momsCombHISQ} for the 0.076 fm lattice as a function of $z_{\rm max}$, while the other two finer lattices have been analyzed in \refcite{Gao:2020ito}, though at NLO level. In the fits, we truncate the moments up to $n^{\rm max}=8$, and fix $z_{\textup{min}}$ to be $2a$ to avoid the most serious discretization effect at the first lattice grid $a$. We vary $z_{\textup{max}}$ to check the stability of the fit and its dependence on the range of $z$. With a fixed factorization scale $\mu$ = 2 GeV, we find that the second moment $\langle x^2 \rangle$ can be extracted at a very short distance $\approx$ 0.2 fm, and is almost independent of $z_{\textup{max}}$. This suggests a good predictive power of the NNLO matching coefficients in the region of $z$ under consideration and that the higher-twist effects or other systematic uncertainty are under control within current statistics. Due to the factorial suppression, the higher moments can only be detected at larger $zP_z$ or $z_{\textup{max}}$ as seen from the figure. To estimate the statistical as well as systematic errors from the fit results, we use the same strategy proposed in \refcite{Gao:2020ito}: for each bootstrap sample, we evaluate the average and standard deviation of observable $A$ from a range of $z_{\textup{max}}$ as ${\textup{Mean}(A)}$ and $\textup{SD}(A)$. Then we can obtain the statistical errors of $\overline{\textup{Mean}(A)}$, and take $\overline{\textup{SD}(A)}$ as the systematic error. The estimates using $z_{\rm max}\in[0.48, 0.72]$ fm are shown as the dark (statistical errors) and light (systematic errors) bands in \fig{momsCombHISQ}. We list the moments up to $\langle x^6 \rangle$ extracted from the three ensembles in \tb{momsHISQ}, while $\langle x^8 \rangle$ is consistent with 0, limited by the statistics, and thus is not shown. As one can see, overall agreement can be observed for different ensembles including the most precise $\langle x^2 \rangle$ with about 5\% accuracy. It is worth mentioning that the $a = 0.04$ fm and 0.06 fm ensembles have an unphysical pion mass of 300 MeV, while the $a = 0.076$ fm one is at the physical point, suggesting the mass dependence of the moments is only mild within current statistics.

\subsection{Continuum estimate of the moments making use of 
the observed weak quark mass dependence}

\begin{figure}
\centering
	\includegraphics[width=0.4\textwidth]{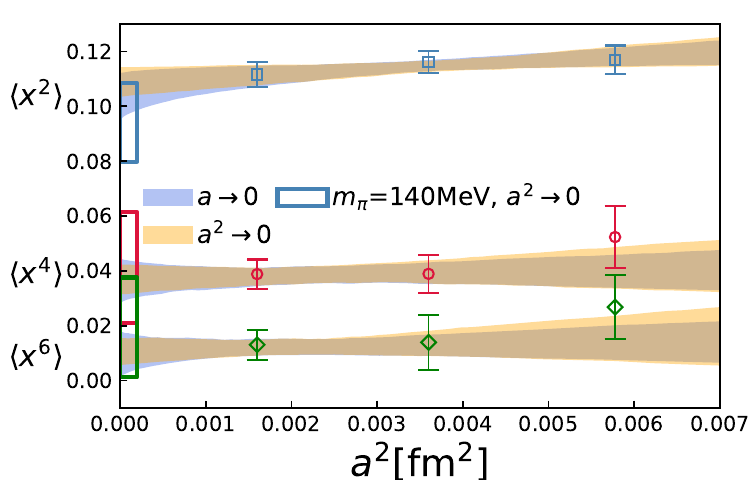}
	\caption{$\langle x^2 \rangle$, $\langle x^4 \rangle$ and $\langle x^6 \rangle$ are shown as a function of $a^2$, for which we fitted the matrix elements $\mathcal{M}(z,P_z,P_z^0)$ with $z\in[2a,0.61]$ fm. The bands are the extrapolation results using \Eq{aExtra} (blue) and \Eq{a2Extra} (orange). The empty boxes are the continuum extrapolation of the physical pion mass ensemble (see the text for more details). \label{fig:contimomstest}}
\end{figure}

\begin{figure}
\centering
	\includegraphics[width=0.4\textwidth]{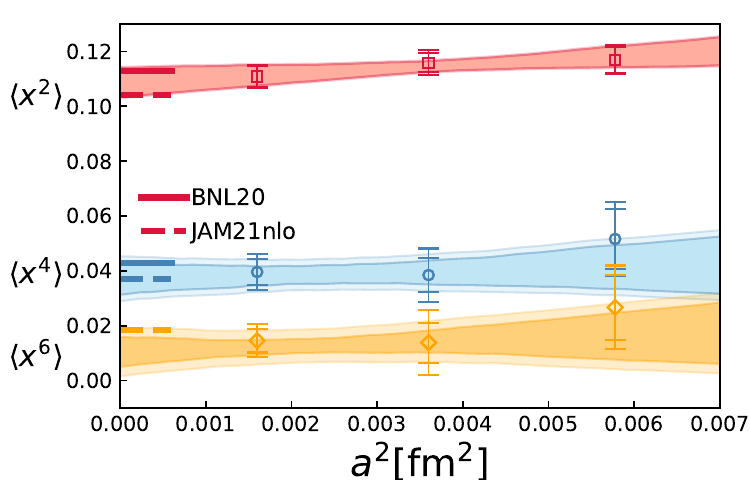}
	\caption{Our estimates for $\langle x^2 \rangle$, $\langle x^4 \rangle$ and $\langle x^6 \rangle$ as functions of $a^2$ are shown for the three ensembles. The bands are from the $a^2$ continuum extrapolation. For comparison, we also show the moments evaluated from the global fit analysis of JAM21nlo~\cite{Barry:2021osv} and our previous estimates with a 300 MeV pion mass using NLO kernels (BNL20)~\cite{Gao:2020ito}.\label{fig:contimom}}
\end{figure}

In \fig{contimomstest}, we show the $\langle x^2 \rangle$, $\langle x^4 \rangle$ and $\langle x^6 \rangle$ 
obtained from fits of $\mathcal{M}(z,P_z,P_z^0)$ in $z\in[2a,0.61]$ fm as a function of $a^2$. Notice that we have combined the data from the physical pion
and the 300 MeV pion together in this plot. As one can see, the mild lattice spacing dependence shows almost 
no pion mass dependence. It is then reasonable to perform continuum extrapolations assuming $a$ or $a^2$ dependence under a justified assumption that we can neglect the 
pion mass dependence. Based on the above observation, we consider two strategies for obtaining the continuum estimate:
\begin{enumerate}
  \item Ignoring the pion mass dependence and performing a continuum extrapolation using the following forms,
  \begin{align}\label{eq:aExtra}
    \langle x^n \rangle_a=\langle x^n \rangle_{a\rightarrow0}+d_na,
  \end{align}
  or
  \begin{align}\label{eq:a2Extra}
    \langle x^n \rangle_{a}=\langle x^n \rangle_{a^2\rightarrow0}+d_na^2.
  \end{align}
  We insert the above formulas into \Eq{ratioTMClat} and perform a joint fit of all the data from the three ensembles instead of directly extrapolating the extracted moments.
  \item We only perform the continuum extrapolation of the two ensembles with a 300 MeV pion mass to obtain the value of $d_n$, and then apply the parameter $d_n$ to the physical pion ensemble ($m_\pi=140$ MeV) to derive the continuum estimate at the physical point. In this procedure, we 
  assume $\langle x^n\rangle_{a^2\to 0}$ could have pion mass dependence, but 
  the values of $d_n$ have negligible pion mass dependence.
\end{enumerate}
The bands in \fig{contimomstest} are derived from the joint fit of all the three ensembles by applying \Eq{aExtra} (blue) and \Eq{a2Extra} (orange) and ignoring the pion mass difference. As one can see, the two bands overlap with each other and pass through the data points with reasonable $\chi^2/d.o.f$ around 1. The pion mass dependence is indeed only mild for the data under consideration as also observed in \refcite{Oehm:2018jvm}. To be conservative, we also apply the second strategy described above.  The extrapolated results are shown as the boxes in \fig{contimomstest}. Limited by the number of ensembles and the statistics, solving both the lattice spacing and mass dependence makes the extrapolation rather unstable and produces large errors bars covering the estimates from the first strategy. We therefore will simply ignore the pion mass dependence.
Considering that the results from \Eq{aExtra} overlap with \Eq{a2Extra} with only a slightly larger error, and the
fact that the Wilson-clover action is $\mathcal{O}(a)$ improved, we will only give the mass independent continuum estimate using the $a^2$ correction of \Eq{a2Extra} in the following analysis.

In \fig{contimom}, we show the moments extracted from the three ensembles with both statistical and systematic errors, estimated by varying $z_{\rm max}\in[0.48,0.72]$ fm. The bands are the continuum estimate using \Eq{a2Extra}. It can be observed that the systematic errors are small compared to the statistical errors, suggesting the small $z_{\rm max}$ dependence for the data under consideration. Our previous results obtained from NLO kernels with a 300 MeV pion (denoted by BNL20~\cite{Gao:2020ito}) are shown for comparison. The good agreement between BNL20 and the new results suggests that the NNLO corrections make only a small difference, and the NLO kernels are mostly sufficient to describe the data evolution with current statistics. The estimated moments are also close to the values obtained from the global fit, JAM21nlo~\cite{Barry:2021osv}.

\section{Pion valence PDF from model dependent fits}\label{sec:modelfit}

\begin{figure}
\centering
	\includegraphics[width=0.4\textwidth]{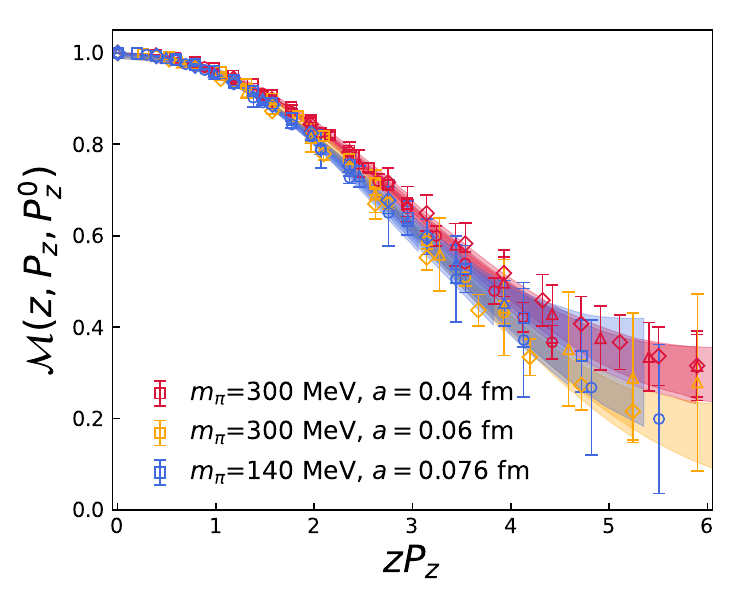}
	\caption{Ratio scheme renormalized matrix elements $\mathcal{M}(z,P_z,P_z^0)$ of $P_z\textgreater P_z^0=0.25$ GeV are shown for the three ensembles, with the bands being the fit results of the Model-$4p$ (c.f. \Eq{modelansatz}) using $z\in [2a, 0.61 {\rm fm}]$.\label{fig:HISQmodelrITD}}
\end{figure}

\begin{figure}
\centering
	\includegraphics[width=0.4\textwidth]{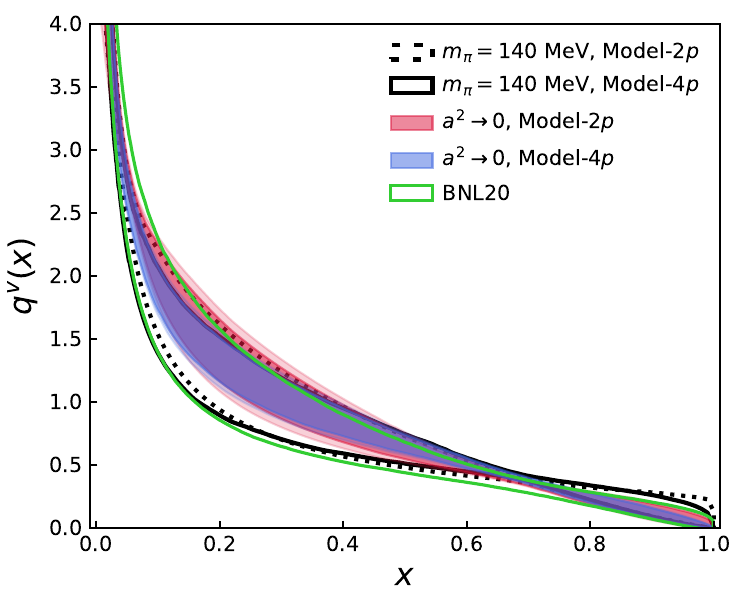}
	\caption{The 2-parameter and 4-parameter fit results from the $m_\pi=140$ MeV ensemble are shown as empty bands. We also show our mass independent continuum estimate as the filled bands using the data with $z_{\rm max}\in [0.48, 0.72]$ fm to estimate the statistical errors (darker bands) and systematic errors (light bands). For comparison, we also show our previous NLO determination (BNL20)~\cite{Gao:2020ito}.~\label{fig:HISQmodelcmp}.
 }
\end{figure}

\begin{table}
\centering
\begin{tabular}{c c c c c}
\hline
\hline
&$\alpha$&$\beta$&$s$&$t$ \\ [1mm]
\hline
Model-$2p$ & -0.45(18)(6) & 0.79(42)(14) & & \\ [1mm]
Model-$4p$ & -0.52(12)(3) & 1.08(33)(8) &-0.34(33)(6) & 1.82(99)(11) \\ [1mm]
\hline
\hline
\end{tabular}
\caption{The model parameters from the joint fit using the three ensembles with $\mathcal{O}(a^2)$ continuum correction at $\mu$ = 2 GeV are shown. The statistical errors are in the first brackets, while the systematic errors are in the second brackets which are estimated by varying  $z_{\rm max}\in[0.48, 0.72]$ fm for the fits.}
\label{tb:model}\end{table}

As discussed in \sec{moments}, our lattice data is only sensitive to the first few moments of the pion valence PDF. Without prior knowledge or constraints on the higher moments, it is impossible to determine the PDFs uniquely. Therefore, as is typical in some global analyses of experiment data, in this section we try two phenomenology inspired model Ans\"atze for the PDF to fit the lattice data,
\begin{align}\label{eq:modelansatz}
\begin{aligned}
q(x;\alpha,\beta) &= \mathcal{N}x^\alpha(1-x)^\beta,\\
q(x;\alpha,\beta, s, t)& = \mathcal{N'}x^\alpha(1-x)^\beta(1+s\sqrt{x}+tx),
\end{aligned}
\end{align}
denoted by Model-$2p$ and Model-$4p$, in which $\mathcal{N}$ and $\mathcal{N'}$ are normalization factors so that $\int_0^1q^v(x)\,dx=1$. The moments of the PDF therefore can be determined from the model parameters, e.g,
\begin{align}
    \langle x^n \rangle_{(\alpha,...)}=\int_{0}^1x^{n}q^v(x;\alpha,...)\,dx .
\end{align}
We then re-express \Eq{ratioTMClat} using the model parameters and minimize,
\begin{align}
\begin{split}
       &\chi^2\\
       &=\sum_{P_z\textgreater P_z^0}^{P_z^{\textup{max}}}\sum^{z_\textup{max}}_{z_\textup{min}}\frac{(\mathcal{M}(z,P_z,P_z^0)-\mathcal{M}_{\textup{model}}(z,P_z,P_z^0;\alpha,...))^2}{\sigma^2(z,P_z,P_z^0)}
\end{split}
\end{align}
where we truncate the OPE formula \Eq{ratioTMClat} up to 20th order ($n^{\rm{max}}=20$), which is much more than sufficient to describe the data. During the fit, the factorization scale was chosen to be $\mu$ = 2 GeV and the correlation between different $P_z$ and $z$ is taken care of by the bootstrap procedure. To apply the leading-twist approximation we need to limit the maximum of $z$, which is chosen to be $z_{\rm max}=0.72$ fm as was done in \sec{moments}. 

We first perform the model fit for the $m_\pi$ = 140 MeV ensemble using the matrix elements with $z\in [2a, 0.61 {\rm fm}]$ shown as the black curves in \fig{HISQmodelcmp}. Then, as with the calculations of the Mellin moments, we perform the mass independent continuum estimate by a joint fit of the three ensembles with the $\mathcal{O}(a^2)$ correction of \Eq{a2Extra} using the ratio-scheme renormalized matrix elements with $P_z\textgreater P_z^0=2\pi n_z^0/(L_sa)$ and $n_z^0=$ 1 for all three ensembles.  The fit result of Model-$4p$ using $z_{\rm{max}}=0.61$ fm is shown in \fig{HISQmodelrITD}. We see that the bands can describe the data well. We vary the $z_{\rm{max}}\in[0.48,0.72]$ fm to estimate the  systematic errors and show the fit results in \tb{model} and the reconstructed PDFs in \fig{HISQmodelcmp} as the red and blue bands. Overall consistency between the two models can be observed in \fig{HISQmodelcmp}. Furthermore, the results overlap with our previous NLO determination (BNL20)~\cite{Gao:2020ito} but has smaller errors since we have an additional data set. 
The good agreement with the previous results again suggests that the use of NNLO kernels did not change the results significanly.

\section{DNN representation of Ioffe-time distribution $Q(\lambda,\mu)$.}\label{sec:DNNITD}

\subsection{The DNN representation}

\begin{figure}
\centering
	\includegraphics[width=0.4\textwidth]{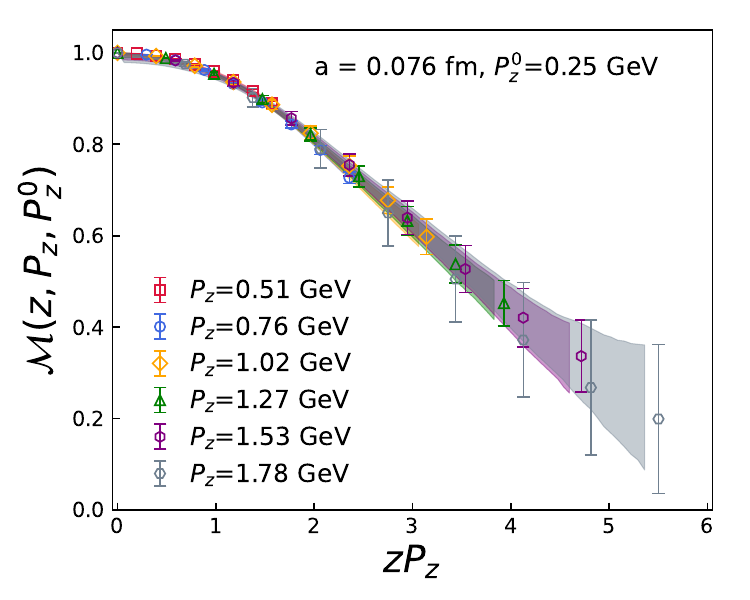}
	\includegraphics[width=0.42\textwidth]{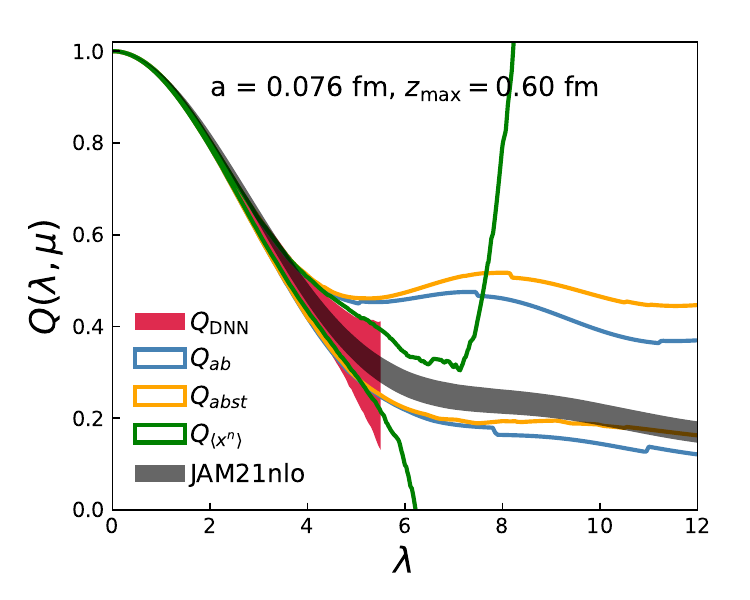}
	\caption{Upper panel: Ratio scheme renormalized matrix elements $\mathcal{M}(z,P_z,P_z^0)$ with $P_z\textgreater P_z^0=0.25$ GeV for the $m_\pi$ = 140 MeV ensemble, where the bands are the fit result from the DNN trained $Q(\lambda,\mu)$. Lower panel: The corresponding DNN represented $Q_{\rm{DNN}}(\lambda, \mu)$ is shown, together with the ones constructed from the moments estimated in \sec{moments} ($Q_{\langle x^n \rangle}$ as green curves) and the model fits in \sec{modelfit} ($Q_{ab}$ and $Q_{abst}$ as blue and yellow curves). All the training and fits in this figure used data on $z\in [2a, 0.61 {\rm fm}]$. The most recent global analysis result from JAM~\cite{Barry:2021osv} (black bands) is also shown for comparison\label{fig:HISQDNNrITD}}
\end{figure}


The short distance factorization formula for
the renormalized matrix element at leading twist (twist-2) can be written as
\begin{align}
h^R(z,P_z,\mu)=\int_{-1}^1d\alpha\, \mathcal{C}(\alpha, \mu^2z^2)\, Q(\alpha\lambda,\mu),
\end{align}
where 
\begin{align}
Q(\lambda,\mu)=\int_{-1}^1 dy\, e^{-iy \lambda} q(y,\mu),
\end{align}
and $\mathcal{C}(\alpha, \mu^2z^2)$ is related to the Wilson coefficients $C_n(\mu^2 z^2)$ \cite{Radyushkin:2017cyf,Orginos:2017kos,Izubuchi:2018srq}. The explicit
form of $\mathcal{C}(\alpha, \mu^2z^2)$ up to NNLO is given in Appendix \ref{app:DNNsetup}.
Therefore, for the ratio scheme matrix element we could write
\begin{align}\label{eq:OPEtw2ITD}
\begin{aligned}
&\mathcal{M}(z,P_z,P_z^0)\\
&=\frac{\int_{-1}^1d\alpha\, \overline{\mathcal{C}}(\alpha, \mu^2z^2)\,Q(\alpha\lambda,\mu)+r(aP_z)^2}{\int_{-1}^1d\alpha\, \overline{\mathcal{C}}(\alpha, \mu^2z^2)\,Q(\alpha\lambda^0,\mu)+r(aP_z^0)^2},
\end{aligned}
\end{align}
where  $\overline{\mathcal{C}}(\alpha, \mu^2z^2) = \mathcal{C}(\alpha, \mu^2z^2)/C_0^{\overline{\textup{MS}}}(\mu^2z^2)$ so that the integral (without the lattice correction $r(aP_z)^2$) in the numerator and denominator is the standard reduced Ioffe-time distribution~\cite{Radyushkin:2017cyf, Orginos:2017kos}. As mentioned in previous sections, the range in  ($\alpha \lambda$) in the above equation depends on the largest $\lambda_\textup{max}=z_\textup{max}P_z^\textup{max}$ achieved from the lattice calculation, which therefore limits the information on the PDFs that the lattice data carries. One can either reconstruct the PDFs $q(y,\mu)$ by a certain model as we did in \sec{modelfit}, or one can directly extract the light-cone Ioffe-time distribution $Q(\lambda,\mu)$. In other words, $Q(\lambda,\mu)$ is the first-principles model-independent observable we can derive from the ratio scheme renormalized matrix elements. For this, however, we need to solve the inverse problem in \Eq{OPEtw2ITD}. So far the deep neural network (DNN) technique is probably the most flexible way to achieve this~\cite{leshno1993multilayer, kratsios2021universal}, which has been used to parametrize the $x$-dependent PDFs from lattice~\cite{Karpie:2019eiq, DelDebbio:2020rgv} and also has been proven effective in other physics inverse problems~\cite{Shi:2021qri,Wang:2021jou}.

In this work, we express the $Q(\lambda,\mu)$ by the DNN function $f_\textup{DNN}(\boldsymbol{\theta};\lambda)$,
\begin{align}
    Q_\text{DNN}(\lambda, \mu) \equiv\frac{f_\textup{DNN}(\boldsymbol{\theta};\lambda)}{f_\textup{DNN}(\boldsymbol{\theta};0)},
\end{align}
where $\boldsymbol{\theta}$ are the DNN parameters. The DNN function $f_\textup{DNN}(\boldsymbol{\theta};\lambda)$ is a multi-step iterative function, constructed layer by layer in composite fashion, to approximate a mapping between two functions in a smooth and unbiased manner. In each layer, the network first performs a linear transformation from the previous layer,
\begin{align}
    z_{i}^{(l)} = b_{i}^{(l)}+\sum_jW_{ij}^{(l)}a_j^{(l-1)}
\end{align}
followed by an element-wise non-linear activation $a_i^{(l)} = \sigma^{(l)}(z_{i}^{(l)})$ then propagate to the next layer ($l+1$). The first layer represents the input variable $\lambda$ and the last layer denotes the corresponding output, $Q_\text{DNN}(\lambda, \mu)$. Here $i = 1,...,n^{(l)}$ and $l=1,...,N$, with $n^{(l)}$ to be the width of the $l$th layer and $N$ the depth of the DNN. The bias $b_{i}^{(l)}$ and the weight $W_{ij}^{(l)}$ are the DNN parameters to be optimized (trained), denoted by $\boldsymbol{\theta}$. We then minimize the loss function,
\begin{align}
\begin{split}
J(\boldsymbol{\theta},r_\text{mod}) 
\equiv
    \frac{\eta}{2} \boldsymbol{\theta}\cdot\boldsymbol{\theta}+\frac{1}{2} \chi^2(\boldsymbol{\theta},r_\text{mod}),
\end{split}
\end{align}
where the first term is to prevent overfitting and make sure the DNN represented function is well behaved and smooth, while the defination and details of $\chi^2$ can be found in \app{DNNsetup}. For the training, we vary $\eta$ from $10^{-1}$ to $10^{-4}$, and tried network structures of size $\{1,16,16,1\}$, $\{1,16,16,16,1\}$ and $\{1,32,32,1\}$ including the input and output layer. We found the results remain unchanged. We therefore chose $\eta=0.001$ and the DNN structure with 4 layers, including the input/output layer, to be $\{1,16,16,1\}$. The exponential linear units ($\textsf{elu}$) were chosen as the action function,
\begin{align}
    \sigma_{\textsf{elu}}(z)=\theta(-z)(e^z-1)+\theta(z)z.
\end{align}
This setup is more than sufficient for the complexity of the data under consideration.

The DNN training process works very much like an interpolation procedure, which is trying to go through the data points as much as possible with many neurons, but smoothly forced by the regularization term $\eta \boldsymbol{\theta}^2$. In other words, the $\chi^2$ of DNN is approaching 0 as much as possible. For our specific task of training the ratio-scheme renormalized matrix elements, part of our data points share the same $\lambda=zP_z$ but different $z$ connected by the perturbative matching kernel $\mathcal{C}(\alpha,\mu^2z^2)$. The minimum of $\chi^2$ is free of the goodness of model, but will depend on how well the kernel can describe the data evolution and usually cannot make it to be zero. In the upper panel of \fig{HISQDNNrITD} we show the $\mathcal{M}(z,P_z,P_z^0)$ obtained from the DNN representation of the light-cone Ioffe time distribution, $Q_\text{DNN}(\lambda,\mu)$. In these training processes we use the ratio-scheme renormalized matrix elements on $z\in [2a$, 0.61 fm]. As one can see, the bands smoothly go through the ratio-scheme data. The results for $Q_\text{DNN}(\lambda,\mu)$ are shown in the lower panel of \fig{HISQDNNrITD}. For comparison, we also reconstruct and show the $\overline{\rm MS}$ ITD from the moments estimated in \sec{moments} (green curves):
\begin{align}
Q_{\langle x^n \rangle}(\lambda,\mu)=\sum_{n=0}\frac{(-i \lambda)^n}{n!}\langle x^n\rangle
\end{align}
and the model fits in \sec{modelfit} (blue and yellow curves):
\begin{align}
\begin{split}
Q_{ab}(\lambda,\mu)&=\int_0^1dx\,e^{-ix\lambda}q^v(x;\alpha,\beta),\\
Q_{abst}(\lambda,\mu)&=\int_0^1dx\,e^{-ix\lambda}q^v(x;\alpha,\beta,s,t).
\end{split}
\end{align}
The ITD from all the methods are consistent with each other within the errors. These observations justify the assertion that our data is only sensitive to the first few moments (up to around $\langle x^6 \rangle$ as discussed in \sec{moments}), though $Q_{\langle x^n \rangle}$ goes out of control rapidly once $\lambda$ is beyond $\lambda_{\textup{max}}=z_{\textup{max}}P_z^{\textup{max}}$ due to the truncation of the OPE formula. The $Q_{ab}(\lambda,\mu)$ and $Q_{abst}(\lambda,\mu)$ are more stable by expressing all the 
Mellin moments with only a few parameters
and modeling the large $\lambda$ behavior beyond $\lambda_{\textup{max}}=z_{\textup{max}}P_z^{\textup{max}}$. The advantage of the DNN is that it does not truncate the matching formula compared to the moments fit, particularly for the case that the moments are not decaying as fast as a function of order $n$, and is much more flexible than the model-based fit. $Q_\text{DNN}(\lambda,\mu)$ with $\lambda_{\textup{max}}=z_{\textup{max}}P_z^{\textup{max}}$ is therefore the most unbiased first-principles result on the ITD that can be obtained from our lattice data in coordinate space. We also show $Q(\lambda,\mu)$ derived from the JAM global analysis results~\cite{Barry:2021osv} in \fig{HISQDNNrITD}, which are in  good agreement with our results.

\subsection{Discussion on the $z_{\rm{max}}$ dependence}

\begin{figure}
\centering
	\includegraphics[width=0.4\textwidth]{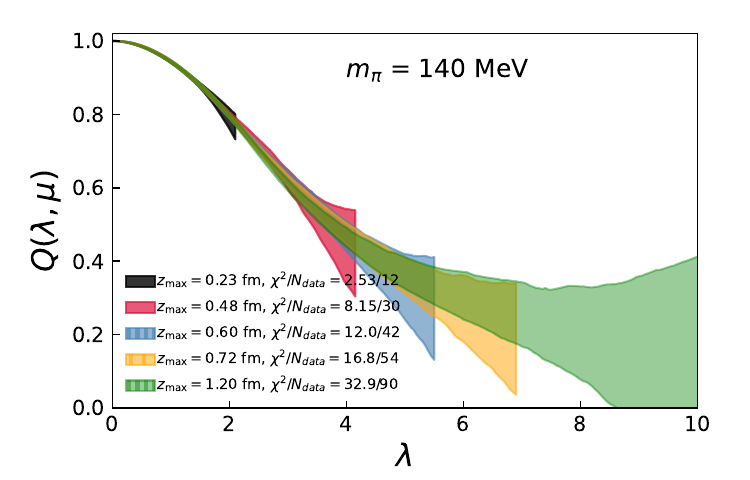}
	\caption{$Q(\lambda,\mu)$ trained from the DNN using data with $z\in [2a, z_{\rm{max}}]$ on the $m_\pi$ = 140 MeV ensemble. The end of the bands depend on $z_{\rm{max}}P_z^{\rm{max}}$.~\label{fig:ITDzmaxLmax10}}
\end{figure}

\begin{figure}
\centering
	\includegraphics[width=0.4\textwidth]{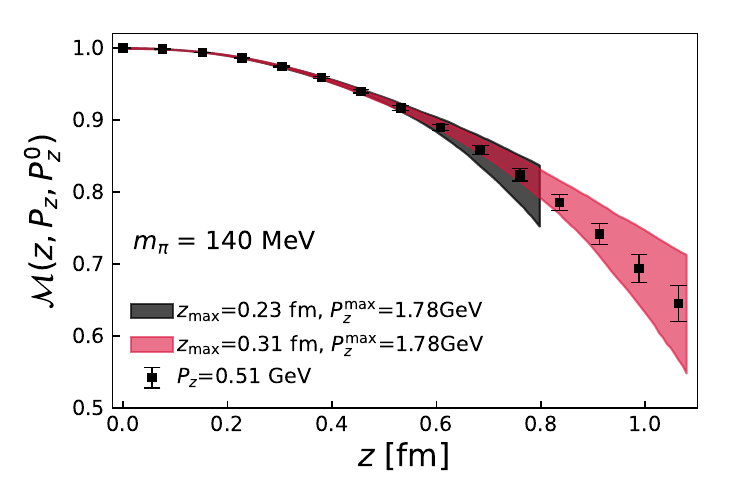}
	\includegraphics[width=0.4\textwidth]{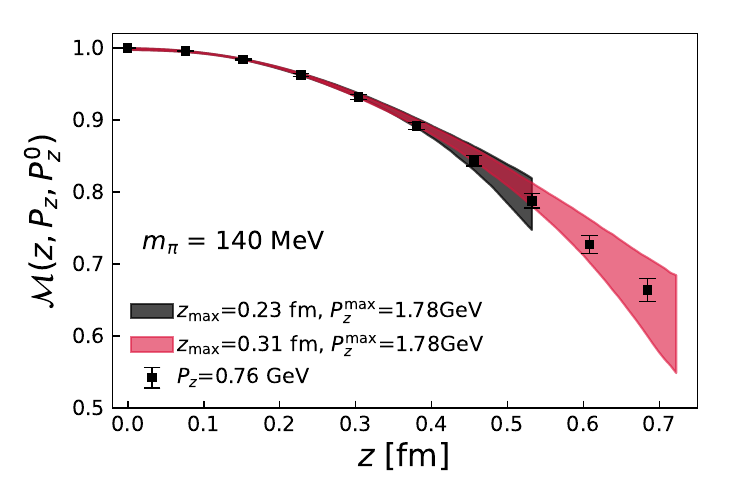}
	\caption{The ratio-scheme renormalized matrix elements $\mathcal{M}(z,P_z,P_z^0)$ with $P_z$ = 0.51 and 0.76 GeV are shown as the data points. The bands are the prediction from the DNN trained results using $P_z$ up to $P_z^{\rm{max}}=1.78$ GeV with a short range of $z_{\rm{max}}$. The end of the bands in $z$ come from $z_{\rm{max}}P_z^{\rm{max}}/P_z$.~\label{fig:rITDSmallPN}}
\end{figure}

As has been mentioned, the short distance factorization scheme suffers from higher-twist corrections proportional to $\mathcal{O}(z^2\Lambda_{\textup{QCD}}^2)$. On the one hand, we want larger $z$ to achieve larger $zP_z$ for a fixed range of $P_z$ values, and thereby,
obtain more information on the PDF from the lattice data.
At the same time we have to ensure that the higher-twist contamination is small. In this section, we discuss how $z_{\rm{max}}$ affects the DNN trained results $Q_\text{DNN}(\lambda,\mu)$.

In \fig{ITDzmaxLmax10}, we show $Q(\lambda,\mu)$ trained from the DNN using data in the range $z\in [2a, z_{\rm{max}}]$ with multiple $z_{\rm{max}}$. The end of the bands depend on the $z_{\rm{max}}P_z^{\rm{max}}$ used in the training process. As one can see, increasing $z_{\rm{max}}$ has little affect on $Q(\lambda,\mu)$ for small $\lambda$ but does decrease the errors. This observation can be explained from the small $\lambda$ region of $Q(\lambda,\mu)$, which is dominated by the lower moments that are well constrained by the precise ratio-scheme lattice data with small $z$. The $\chi^2/N_{\rm{data}}$ are also shown in \fig{ITDzmaxLmax10}. As we mentioned, the matching kernel $\overline{\mathcal{C}}(\alpha, \mu^2z^2)$ is supposed to connect the space-like matrix elements with different $z$ perturbatively. Since the DNN is a flexible parametrization that could pass through all the data points smoothly as much as possible, the non-zero $\chi^2/N_{\rm{data}}$ can only come from the fact that the central values of the lattice data corresponding to the same $zP_z$ but different $z$ are slightly different from the expectation from perturbative evolution. Therefore the small $\chi^2/N_{\rm{data}}$ indicates that the kernel describes the evolution of the matrix elements as a function of $z$ well within the statistical errors. 
We also see that the $Q(\lambda,\mu)$ obtained with $z_{max}=1.2$ fm agrees well with the results obtained
with smaller $z_{max}$ within errors. Thus it is natural to ask how
large $z_{max}$ in the DNN analysis could be. 
To check this we determine $Q(\lambda,\mu)$ with very small $z_{\rm{max}}$ but large $P_z$, and then predict the ratio scheme matrix elements for smaller $P_z$ but larger $z$ through the matching. This prediction is then compared to the corresponding lattice result. In \fig{rITDSmallPN}, we show the ratio-scheme matrix elements $\mathcal{M}(z,P_z,P_z^0)$ of $P_z$ = 0.51 GeV (upper panel) and 0.76 GeV (lower panel) as the black data points, and the bands are predicted from the DNN trained results using $P_z$ up to $P_z^{\rm max}=1.78$ GeV. We use the data in the range $z\in [2a, z_{\rm{max}}]$ with short distances $z_{\rm{max}}$ = 0.23 (black) and 0.31 (red) fm. It is interesting to observe that the predicted bands from short distances are consistent with the matrix elements at relatively large distances, suggesting that the matching kernel works well up to $z\sim$ 0.8 fm from the prediction of $z_{\rm{max}}$ = 0.23 fm, and up to $z\sim$ 1 fm from the prediction of $z_{\rm{max}}$ = 0.31 fm within our current statistics. This observation is consistent with what we observed in \sec{moments} for the moments extraction and seems to support the argument that the ratio-scheme indeed reduces the higher-twist effect $\mathcal{O}(z^2\Lambda_{\rm QCD}^2)$ by the cancellation between the numerator and denominator, even though it is naively not expected that the leading-twist OPE can approximately work up to 1 fm. For the following analysis, we conservatively use $z$ only up to 0.72 fm, and vary $z_{\rm{max}}$ between 0.48 and 0.72 fm to estimate the systematic errors.

\subsection{Discussion on the perturbative order dependence}

\begin{figure}
\centering
	\includegraphics[width=0.4\textwidth]{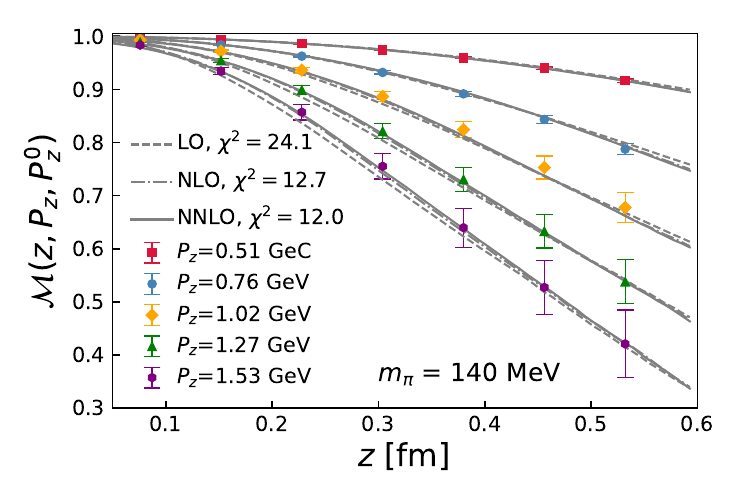}
	\includegraphics[width=0.4\textwidth]{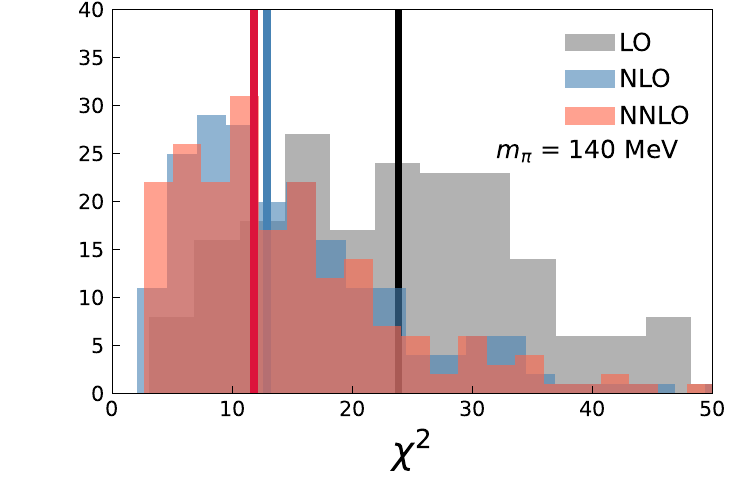}
\caption{Upper panel: $\mathcal{M}(z,P_z,P_z^0)$ reconstructed from the DNN trained $Q(\lambda,\mu)$ are shown for the $m_\pi$ = 140 MeV ensemble using LO, NLO and NNLO matching kernels. The central values for the $\chi^2$ of the bootstrap samples are shown in the legends. The fits use $z\in [2a, z_{\textup{max}}]$ with $z_{\textup{max}}$ = 0.61 fm. Lower panel: the corresponding distribution for the $\chi^2$ of the bootstrap samples with the vertical lines being the median values. \label{fig:DNNOrderCmpHISQa076}}
\end{figure}


In this subsection, we explore the perturbative order dependence through the DNN approach, which does not need a truncation of the OPE matching formula.
In \fig{DNNOrderCmpHISQa076}, the ratio-scheme matrix elements together with the curves reconstructed from the central values of the DNN trained results using LO, NLO and NNLO matching kernels are shown in the upper panel. It is obvious that the LO matching kernel shows the worst performance among the curves, which only marginally passes within the error bars of the data points. Beyond LO, the curves tend to pass through the central values of the data with much smaller $\chi^2$, though the difference between NNLO and NLO is only mild. This
indicates that the leading-twist approximation with fixed-order perturbative kernels works better than LO for the data under consideration, while with current statistics the systematic improvement from NLO to NNLO is small. The distributions for the $\chi^2$ of the bootstrap samples are shown in the lower panel of \fig{DNNOrderCmpHISQa076} where the $\chi^2$ for the LO fit is significantly larger than the other two cases.
\subsection{Extrapolation of the $Q_{\rm{DNN}}(\lambda, \mu)$ and F.T. to the PDFs}

\begin{figure}
\centering
	\includegraphics[width=0.4\textwidth]{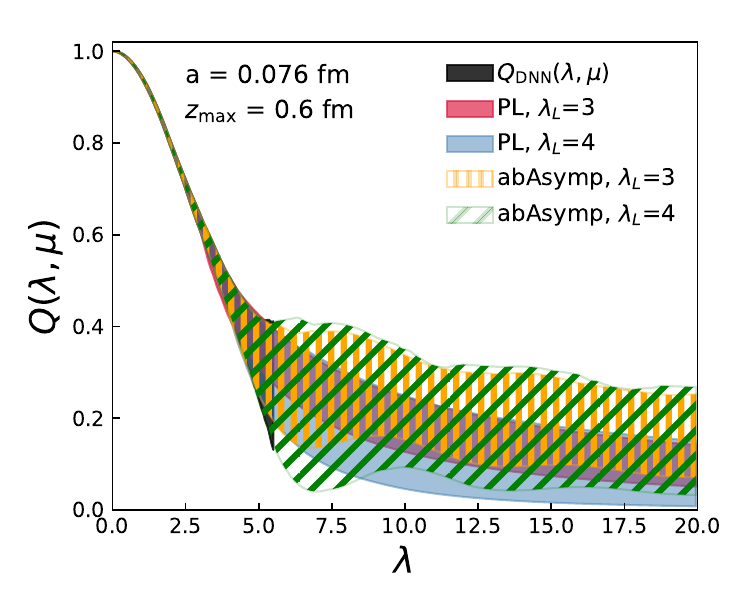}
	\includegraphics[width=0.4\textwidth]{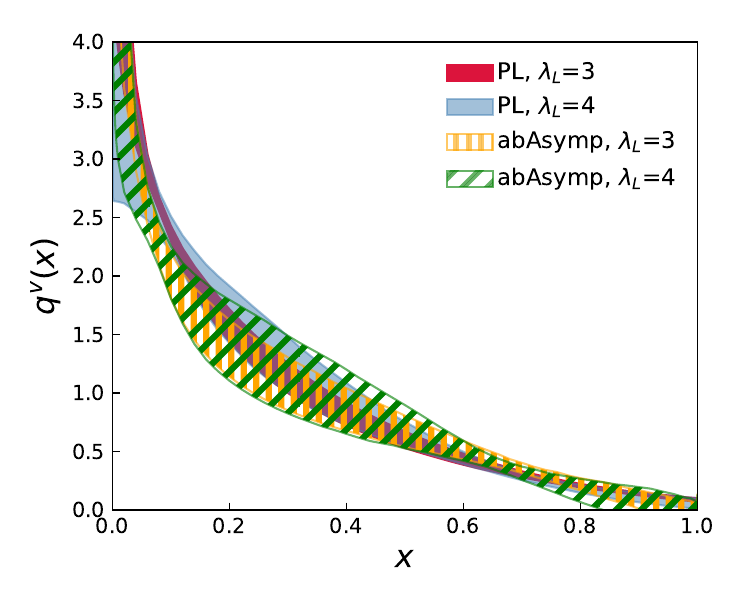}
	\caption{Upper panel: the DNN trained $Q_{\rm{DNN}}(\lambda, \mu)$ including the extrapolations using either the power law (denoted by PL) model or the two-parameter asymptotic (denoted by abAsymp) model. Lower panel: the F.T. of the extrapolated ITD are shown. \label{fig:ITDExtraZmax060}}
\end{figure}


The $Q_{\rm{DNN}}(\lambda, \mu)$ trained from the DNN is the most unbiased and first-principles physical quantity we extracted from the ratio-scheme matrix elements. It is more flexible than the truncated moments extraction (c.f. \sec{moments}) and the model-based fits (c.f. \sec{modelfit}). However, usually one is more interested in the PDF $q(x,\mu)$. The $Q_{\rm{DNN}}(\lambda, \mu)$ we obtained is limited by the lattice data up to $\lambda_{\rm{max}}=z_{\rm{max}}P_z^{\rm{max}}$, so that we need to extrapolate the large $\lambda$ behavior to infinity to do the Fourier transform.

Inspired by Regge theory, the asymptotic behavior of $Q(\lambda, \mu)$ is dominated by the power law (PL) form,
\begin{align}
    Q_{\rm{PL}}(\lambda, \mu) = \frac{A}{|\lambda|^{\alpha+1}}.
\end{align}
Limited by the $\lambda_{\max}$ we have, the sub-leading contribution could also be significant, therefore this extrapolation has systematic errors from the parameterization. We fit the tail of $Q_{\rm{DNN}}(\lambda, \mu)$ with $\lambda\in[\lambda_L, \lambda_{\rm max}]$ and vary $\lambda_L$ to see the $\lambda_L$ dependence of the extrapolation results. We also impose the constraint $A>0$ during the extrapolation to ensure the extrapolated ITD is positive and decreasing in $\lambda$. Inspired by the asymptotic behavior of the Fourier transform of the two-parameter model $\mathcal{N}x^\alpha(1-x)^\beta$, we also consider the form~\cite{Ji:2020brr, Gao:2021dbh},
\begin{align}
\begin{split}
    Q_{\rm{abAsymp}}(\lambda, \mu)=\Re[ A(\frac{\Gamma(1+\alpha)}{(-i|\lambda|)^{\alpha+1}}+e^{i\lambda}\frac{\Gamma(1+\beta)}{(i|\lambda|)^{\beta+1}})],
\end{split}
\end{align}
where we have three parameters and we impose the constraint $A>0$ and $\beta > 0$. We then combine the DNN trained results and the extrapolated results together as,
\begin{align}
Q(\lambda, \mu)=\left\{
\begin{aligned}
Q_{\rm{DNN}}(\lambda, \mu)   \quad &\mbox{for} &\lambda &\leq&\lambda_L\\
Q_{\rm{Extra}}(\lambda, \mu) \quad &\mbox{for} &\lambda &>&\lambda_L.
\end{aligned}
\right.
\end{align}
The extrapolated results are shown in the upper panel of~\fig{ITDExtraZmax060}, where $Q_{\rm{DNN}}(\lambda, \mu)$ is trained using $z\in[2a,0.61$ fm] for the $m_\pi$ = 140 MeV lattice. We fit the extrapolation form using $\lambda\in[\lambda_L, \lambda_{\rm max}]$ with $\lambda_L$ = 3 and 4. 
As one can see, the extrapolation results using different $\lambda_L$ overlap with each other. The two-parameter model (abAsymp) extrapolated results always show wider error bands compared to the power law model (PL) since one more parameter is used in the fit form. We then perform the Fourier transform by combining the discrete Fourier transform and the analytical integral,
\begin{align}
\begin{split}
q(x,\mu)&\equiv\sum_{0}^{\lambda_{\rm max}}2\frac{\Delta\lambda}{2\pi}Q_{\rm{DNN}}(\lambda, \mu)\cos(x\lambda)\\
&+2\int_{\lambda_{\rm max}}^{\infty}\frac{d\lambda}{2\pi}Q_{\rm{Extra}}(\lambda, \mu)\cos(x\lambda),
\end{split}
\end{align}
in which $\Delta\lambda$ is the step spacing of $\lambda$ that was used when we trained from the DNN. The results of the Fourier transform are shown in the lower panels of \fig{ITDExtraZmax060}. The bands from different extrapolation models and different choices of $\lambda_L$ overlap with each other and the discrepancy among them mainly exists in the small-$x$ region.

\subsection{Continuum estimate of the DNN represented ITD}

\begin{figure}
\centering
	\includegraphics[width=0.4\textwidth]{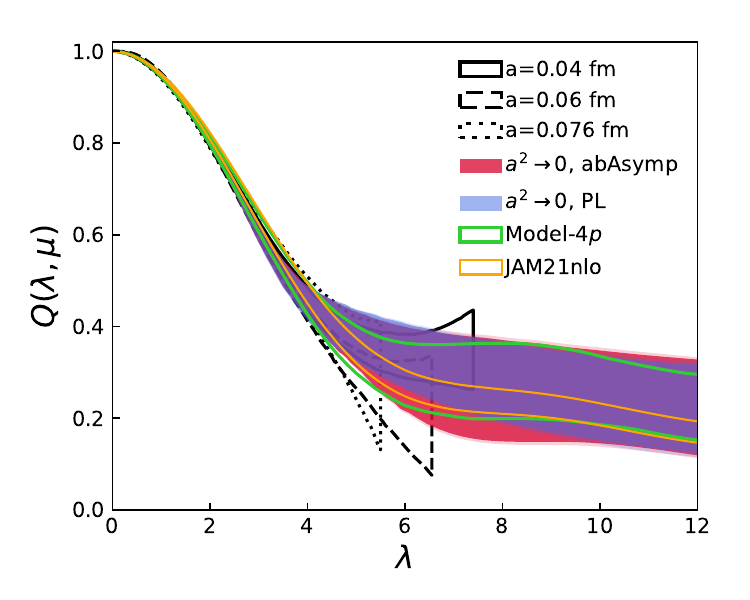}
	\caption{The $Q(\lambda,\mu)$ reconstructed from the DNN using $z\in$ [$2a$, 0.61 fm] for the three ensembles are shown as the empty bands, while the orange band is the continuum estimate with $\mathcal{O}(a^2)$ corrections. We also show the ITD reconstructed from JAM21nlo~\cite{Barry:2021osv} for comparison.\label{fig:ITDContiTest}}
\end{figure}

\begin{figure}
\centering
	\includegraphics[width=0.4\textwidth]{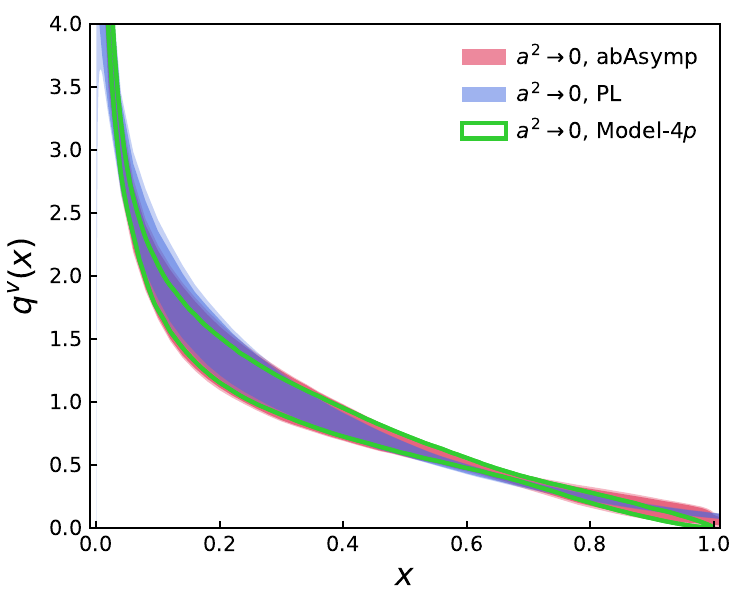}
	\caption{The results for the pion valence distribution from the Fourier transform of the mass independent continuum estimate of the DNN based $Q(\lambda,\mu)$ using either the power-law (PL) or two-parameter asymptotic (abAsymp) extrapolations are shown. We also show the results from the Model-$4p$ fit for comparison.\label{fig:contimodel}}
\end{figure}

We summarize the $Q(\lambda,\mu)$ from the three HISQ ensembles using data with $z\in[2a,0.61 {\rm fm}$] in \fig{ITDContiTest}. As expected from the previous analyses of the moments and model fits, the estimates from the three ensembles are close to each other. We again consider the continuum extrapolation only of order $\mathcal{O}(a^2)$ as,
\begin{align}
    Q_a(\lambda,\mu)=Q_{a^2\rightarrow0}(\lambda,\mu)+g(\lambda)a^2 ,
\end{align}
where $g(\lambda)$ is an independent fit parameter for each $\lambda$, from which we then perform the extrapolation using the power-law (PL) and the two-parameter asymptotic (abAsymp) models with $\lambda_L$ varying from 3 to 4.5.  As before, 
we ignore any weak quark mass dependence in the data sets. To stabilize the two-parameter asymptotic (abAsymp) extrapolation, we impose a prior on the exponent $\alpha$, which dominates the decay rate as a function of $\lambda$, using the corresponding value from the power-law fit sample by sample. The extrapolated results are shown as the red and blue bands in \fig{ITDContiTest}. We compare our continuum estimated ITD to the ones from our model fit Model-$4p$, and the global analysis of JAM21nlo~\cite{Barry:2021osv}. It can be seen that our results show good agreement with JAM21nlo in the small-$\lambda$ region, while it only covers the JAM2nlo at large $\lambda$ due to the large errors. In \fig{contimodel}, we show the results for the pion valence distribution from the Fourier transform of the extrapolated ITD, where we vary $z_{\rm max}\in$ [0.6, 0.72] fm to estimate the statistical and systematic errors. Overall agreement can be observed compared to our model based determination Model-$4p$, while as expected the DNN based PDF, which is more flexible, shows larger errors than our model based determination. Since the power-law extrapolation (blue band) does not ensure that the PDFs are only defined in $x\in$ [0, 1], it does not vanish at $x=$ 1. This can be considered as a systematic error from the extrapolation model which parametrizes the unknown large-$\lambda$ behavior. With one more parameter, the results from the two-parameter asymptotic (red band) extrapolation show larger statistical errors but are consistent with 0 at $x$ = 1.

\section{Pion valence PDF from the quasi-PDF approach with hybrid renormalization}\label{sec:hybrid}

\begin{figure}
\centering
	\includegraphics[width=0.4\textwidth]{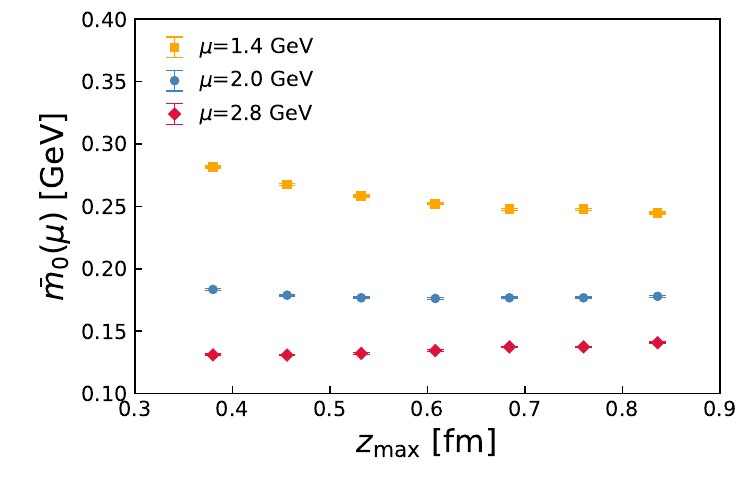}
	\includegraphics[width=0.4\textwidth]{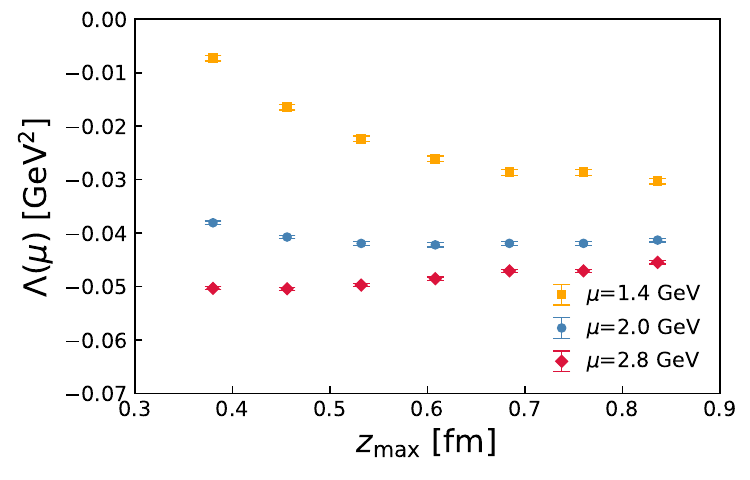}
	\caption{The results for $\bar{m}_0(\mu)$ and $\Lambda(\mu)$ from fits of $\tilde{R}(z,z_S,P_z=0)$ with $z\in[z_S+a,z_{\rm max}]$ to \Eq{ansatz} are shown. We choose the scale $\mu=$ 2 GeV, and vary it by a factor of $\sqrt{2}$. \label{fig:hybrid-2p}}
\end{figure}
Another way to obtain the valence pion PDF from the matrix elements calculated in this study is to use the quasi-PDF approach of LaMET. Very recently, using the NNLO quasi-PDF matching with the novel hybrid renormalization
scheme~\cite{Ji:2020brr}, the valence pion PDF has been calculated with controlled systematic errors
in the range $0.03\lesssim x \lesssim 0.8$~\cite{Gao:2021dbh}. In that study, two lattice spacings, $a=0.04$ fm and $a=0.06$ fm were used, and
the discretization effects turned out to be small. On the other hand, the valence quark masses were
larger than the physical ones, corresponding to a pion mass of $300$ MeV. In this section we
present a calculation of the valence pion PDF within the quasi-PDF approach using NNLO matching and the
hybrid renormalization scheme for physical quark masses ($M_{\pi}=140$ MeV) and $a=0.076$ fm.
Our analysis  closely follows the strategy outlined in Ref.  \cite{Gao:2021dbh}.

The hybrid renormalization scheme is defined by a combination of the ratio scheme at short distances and the
explicit subtraction of the self energy divergence of the Wilson line at large distances:
\begin{align}\label{eq:hdef}
    \tilde{R}(z,z_S,P_z) = 
   \begin{cases}
   \frac{h^B(z,P_z,a)}{h^B(z,0,a)},&{z\leq z_S}\\
   \frac{h^B(z,P_z,a)}{h^B(z_S,0,a)}e^{\delta m(a) |z-z_S|},&{z> z_S}.
   \end{cases}
\end{align}
Here $\delta m(a)$ is the parameter containing the self-energy divergence of the Wilson line.
It is
determined from the analysis of the free energy of a static quark~\cite{Bazavov:2018wmo, Petreczky:2021mef}.
In this work, we choose $z_S$ = 0.228 fm for the $a = 0.076$ fm ensemble.
We need to connect $h^R$ in the hybrid scheme to $h^R$ in the $\overline{\rm MS}$ scheme, and to do this we need
to introduce a parameter, $\bar m_0$, that connects the renormalon ambiguity in the renormalization of
the Wilson line in the lattice subtraction scheme and $\overline{\rm MS}$ scheme \cite{Gao:2021dbh}. As in Ref.~\cite{Gao:2021dbh}, to obtain $\bar m_0$ we fit our results for $\tilde{R}(z,z_S,P_z=0)$ to the following
Ansatz:
\begin{align}\label{eq:ansatz}
\begin{aligned}
    &\tilde{R}(z,z_S,P_z=0) \\
    &= e^{-\bar{m}_0(\mu)(z-z_S)}  \frac{C_0^{\rm NNLO}(z^2\mu^2)  + \Lambda(\mu) z^2}{C_0^{\rm NNLO}(z_S^2\mu^2)  + \Lambda(\mu) z_S^2},
\end{aligned}
\end{align}
with another parameter $\Lambda(\mu)$ that approximates the higher-twist contributions. 
We perform the above 2-parameter fit with $z\in[z_S+a,z_{\rm max}]$. At fixed order, $\bar{m}_0(\mu)$ and $\Lambda(\mu)$ are both scale dependent. We choose $\mu=$ 2 GeV
as the central value of the renormaliation scale and vary it by a factor of $\sqrt{2}$ to estimate the perturbative uncertainty. The fit results are shown in \fig{hybrid-2p}. 
As one can see from the figure, the resulting values of $\bar{m}_0(\mu)$ and $\Lambda(\mu)$ 
show only mild dependence on $z_{max}$ for $\mu \ge 2$ GeV. For the lowest value
of the renormalization scale, $\mu=1.4$ GeV, on the other hand, we see significant dependence
on $z_{max}$. This implies that for such low values of $\mu$, the perturbative expression
for $C_0(\mu^2 z^2)$ is not very reliable because $\alpha_s$ is quite large, and the
Ansatz given by Eq. (\ref{eq:ansatz}) cannot describe the data well. We also see
that there is a significant scale dependence for the value of $\bar m_0(\mu)$ and
$\Lambda(\mu)$, which in turn translates into an uncertainty for the $z$-dependence of
$h^R(z,z_S,P_z)$. This uncertainty will get smaller as $P_z$ increases,
and disappears in the $P_z \rightarrow \infty$ limit. For our final estimates of
$\bar m_0$ and $\Lambda$ we chose $z_{max}=0.456$ fm.

\begin{figure}
\centering
	\includegraphics[width=0.45\textwidth]{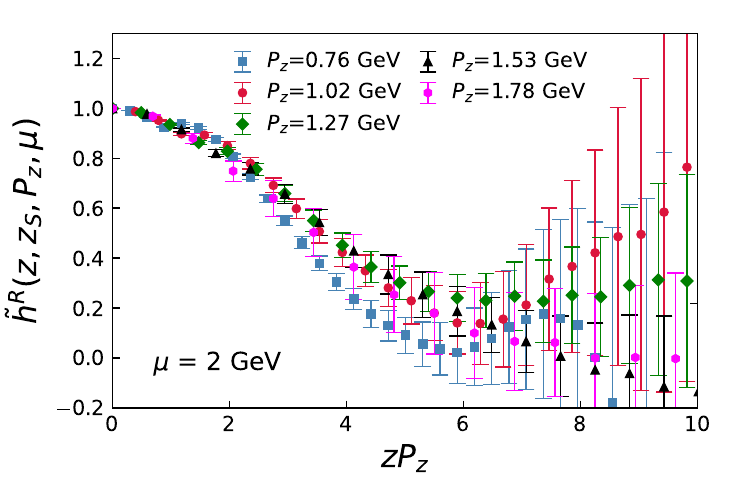}
	\caption{The hybrid-scheme renormalized matrix elements $\tilde h^R(z, z_S,P_z,\mu)$ as a function of $zP_z$ at $\mu=2$ GeV for the $m_\pi=140$ MeV ensemble are shown.\label{fig:hybridmx}}
\end{figure}

\begin{figure}
\centering
	\includegraphics[width=0.45\textwidth]{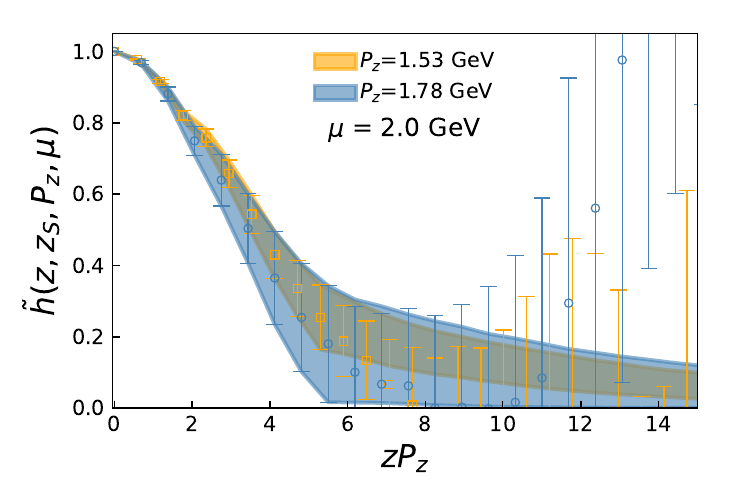}
	\includegraphics[width=0.45\textwidth]{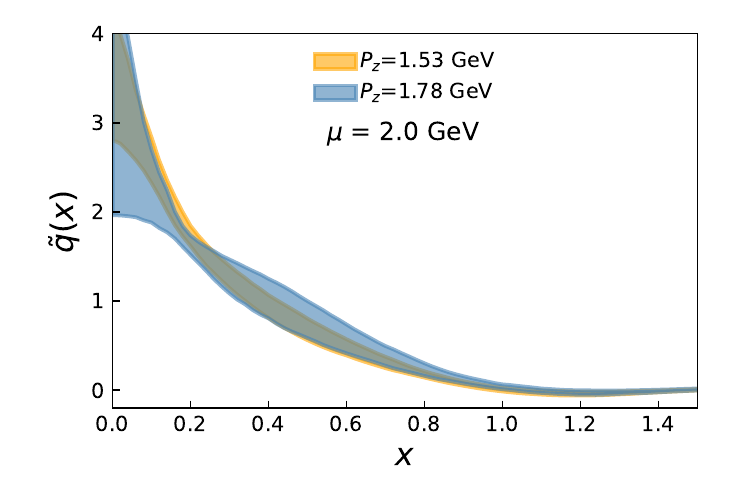}
	\caption{The extrapolated matrix elements $\tilde h^R(z, z_S,P_z,\mu)$ as a function of $zP_z$ at $\mu=2$ GeV are shown in the upper panel for cases $P_z=$ 1.53 GeV and 1.78 GeV. The corresponding quasi-PDFs after a Fourier transform are shown in the lower panel.\label{fig:hybridextra}}
\end{figure}

\begin{figure}
\centering
    \includegraphics[width=0.4\textwidth]{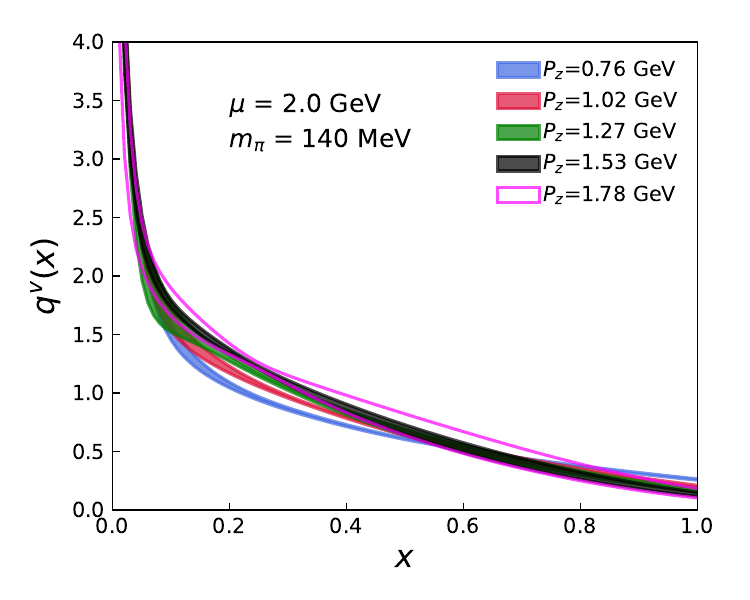}
	\includegraphics[width=0.4\textwidth]{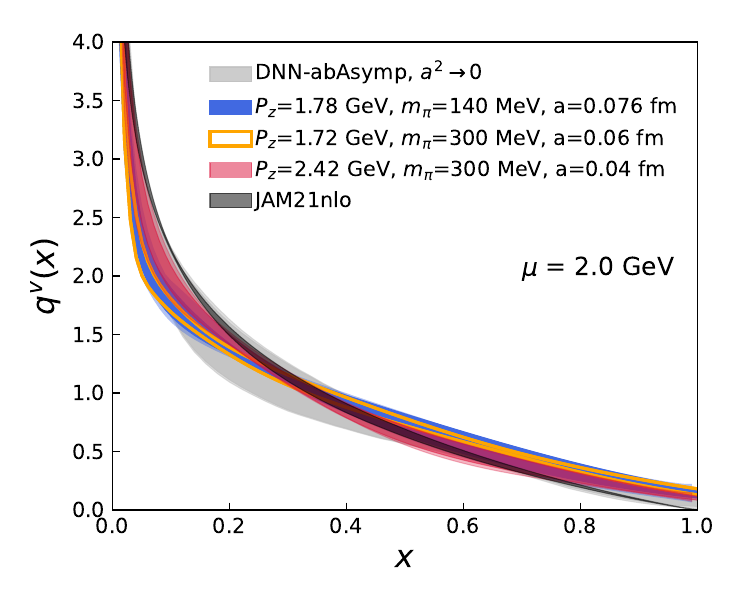}
	\caption{Upper panel: The pion valence distribution $q^v(x)$ obtained from different $P_z$ for the $m_\pi$ = 140 MeV ensemble using NNLO matching is shown. Lower panel: The pion valence distribution $q^v(x)$ obtained from $P_z$ = 1.78 GeV (blue) on the $m_\pi$ = 140 MeV lattice, $P_z$ = 1.72 GeV (orange) and $P_z$ = 2.42 GeV (red) on the $m_\pi$ = 300 MeV lattices are shown. The darker bands come from the statistical errors while the weaker bands are the systematic errors from scale variation. We also show our best determination from the DNN for comparison.\label{fig:hybridPDFfinal}}
\end{figure}

Following Ref. \cite{Gao:2021dbh}, we calculate the renormalized matrix element in the ratio
scheme as
\begin{align}\label{eq:hybren}
\begin{aligned}
   &\tilde h^R(z, z_S,P_z,\mu) \\
   &=\begin{cases}
   N\frac{h^B(z,P_z,a)}{h^B(z,0,a)} {C_0(z^2\mu^2) \!+\! \Lambda z^2\over C_0( z^2\mu^2)},&{z\leq z_S}\\
   N\frac{h^B(z,P_z,a)}{h^B(z_S,0,a)} {C_0(z^2_S\mu^2) \!+\! \Lambda z^2_S \over C_0( z^2_S\mu^2) }e^{\delta m'(z - z_S)},&{z> z_S}.
   \end{cases}
\end{aligned}
\end{align}
where $N=h^B(0,0,a)/h^B(0,P_z,a)$ and $\delta m'=\delta m+\bar m_0$.
The above expression for $\tilde h^R(z,z_S,P_z)$ is equivalent to Eq.~(\ref{eq:hdef}) if lattice
artifacts and higher twist contributions can be neglected. The factor $N$ ensures that the
renormalized matrix element is equal to one at $z=0$ and removes the lattice artifacts 
proportional to powers of $a P_z$. The factor $(C_0(z^2\mu^2+\Lambda z^2)/C_0( z^2\mu^2)$
was introduced to remove the leading higher twist contribution that enters through $h^B(z,0,a)$.
Because of this factor and $\bar{m}_0$, the calculated renormalized matrix element in the ratio scheme also
depends on $\mu$ at any given order of perturbation theory used to evalulate $C_0(\mu^2 z^2)$.
At infinite order, this $\mu$ dependence will disappear. The renormalized matrix elements at $\mu$ = 2 GeV are shown in \fig{hybridmx}. One can observe that $\tilde h^R$ tends to saturate when $P_z\gtrsim 1$ GeV, though the statistical errors are large. 


Since the matrix elements at large $z$ are very noisy and 
becomes unreliable, we shall first perform an extrapolation, and then do the Fourier transform to obtain the quasi-PDF $\tilde q(x)$. For the extrapolation we use $\tilde h^R$ in the range $[z_{\rm min}, z_{\rm max}]$ with $z_{\rm min}$ being the first $z$ where $\tilde h^R(z=z_{\rm min})<0.2$, and $z_{\rm max}$ being either the last $z$, 
where $\tilde h^R(z_{z=\rm max})>0$ or the limit of half the lattice size (32$a$), i.e. $32a$. Since the spatial correlators of the equal-time matrix elements decay exponentially at large $z$ (with the exception of zero modes), we chose the model combining the exponential and power decay as discussed in detail in \refcite{Gao:2021dbh},
\begin{align}
    A {e^{-m_{\rm eff}|z|}\over |\lambda|^d},
\end{align}
with $\lambda=zP_z$, and the constraints $A,d>0$ and $m_{\rm eff}>0.1$ GeV. A continuity condition is also imposed at the point connecting the data and the extrapolation function. The extrapolated matrix elements for $P_z$ = 1.53 GeV and 1.78 GeV are shown in the upper panel of \fig{hybridextra}. Then we perform the Fourier transform (F.T.) by combining the discrete F.T. of the lattice data for $z<z_{\rm min}$ and the continuous F.T. of 
the extrapolation function for $z\geq z_{\rm min}$. The corresponding
results are shown in the lower panel of \fig{hybridmx}. The extrapolation has very weak impact on the moderate-to-large $x$ region, but introduces large uncertainty at small $x$. Nevertheless, after perturbative matching the latter is beyond the region of $x$ where we have systematic control over~\cite{Gao:2021dbh}.

Next, we match the qPDF $\tilde q^v(x,\lambda_S,P_z,\mu)$ to the pion valence distribution $q^v(x,\mu)$ in the $\overline{\rm MS}$ scheme through LaMET~\cite{Xiong:2013bka,Ma:2014jla,Izubuchi:2018srq,Ji:2020brr} using NNLO kernels~\cite{Gao:2021dbh,Chen:2020ody,Li:2020xml}:
\begin{align}\label{eq:fact}
q^v(x, \mu)&= \int_{-\infty}^{\infty} \frac{dy}{|y|} \ C^{-1}\!\left(\frac{x}{y}, \frac{\mu}{yP_z},|y|\lambda_S\right) \tilde q^v(y,\lambda_S,P_z,\mu) \nonumber\\
&\qquad + {\cal O}\Big(\frac{\Lambda_{\text{QCD}}^2}{(xP_z)^2},\frac{\Lambda_{\text{QCD}}^2}{((1-x)P_z)^2}\Big)\,,
\end{align}
where $\lambda_S=z_SP_z$ and $z_S=0.228$ fm. Then we will directly derive the PDF with $P_z$-controlled power corrections for the middle range of $x$. In the upper panel of  \fig{hybridPDFfinal}, we show the PDF obtained from matching with $P_z\in[0.76,1.78]$ GeV. 
We see a significant dependence of the valence pion PDF on $P_z$. However, as $P_z$ increases, this dependence
diminishes, and for the largest three $P_z$ the resulting valence pion PDFs agree within the estimated
errors. The value of $q^v$ for $x \simeq 1$ also decreases with increasing $P_z$.

Finally, we show $P_z$ = 1.78 GeV (blue) for the $m_\pi$ = 140 MeV lattice and $P_z$ = 1.72 GeV (orange) and 2.42 GeV (red) for the $m_\pi$ = 300 MeV lattices from \refcite{Gao:2021dbh} in \fig{hybridPDFfinal}, with the darker bands being the statistical errors and the lighter bands being the systematic errors from scale variation. It can be seen that the systematic errors estimated from scale variation are small, even though $\bar{m}_0(\mu)$ shows sizable $\mu$ dependence. That is because the renormalon effect, which depends on $z$, contributes less for larger momentum and is supposed to disappear for infinite momentum. The results from similar momentum, such as 1.78 GeV and 1.72 GeV, basically overlap with each other within the statistical errors, implying both the lattice spacing and mass dependence are small (more details can be found in \app{qPDFmass}), but differ from the one with larger momentum $P_z$ = 2.42 GeV. The the best determination from the DNN in the continuum limit and the JAM21nlo~\cite{Barry:2021osv} results are also shown for comparison. Though all three LaMET results show some agreement with JAM21nlo in the middle-$x$ region, the one with the highest momentum ($P_z$ = 2.42 GeV) overlaps the best.

\section{Conclusion}\label{sec:conclusion}

\begin{figure}[tp!]
\centering
	\includegraphics[width=0.35\textwidth]{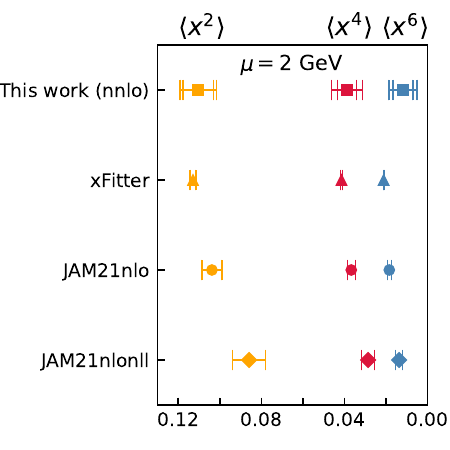}
	\caption{$\langle x^n \rangle$ with $n = 2,~4,~6$ from the mass independent continuum estimate using NNLO matching at $\mu$ = 2 GeV are shown in the first row. The statistical errors are in the first brackets, while the systematic errors are in the second brackets which are estimated by varying  $z_{\rm max}\in[0.48, 0.72]$ fm for the fits. For comparison we also show the moments evaluated from global analyses of xFitter~\cite{Novikov:2020snp}, JAM21nlo and JAM21nlonll double-Mellin~\cite{Barry:2021osv}.\label{fig:momscmp}}
\end{figure}

\begin{figure}[tb!]
\centering
	\includegraphics[width=0.45\textwidth]{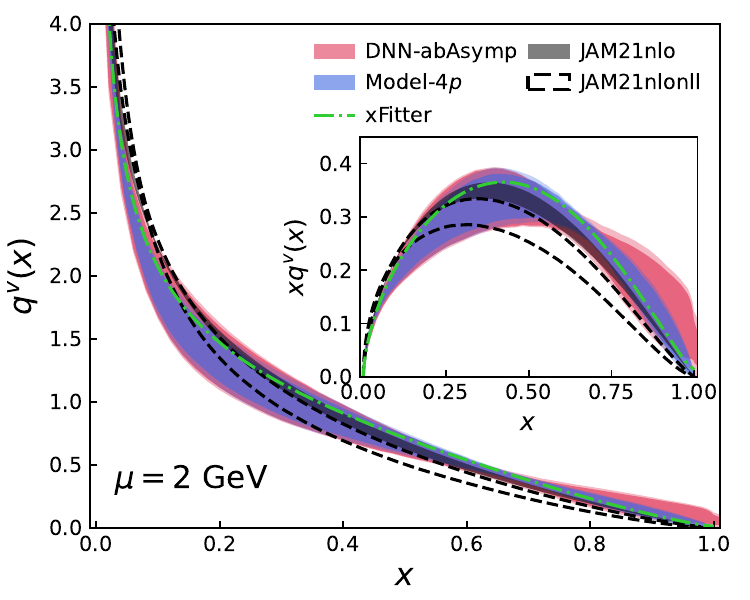}
	\caption{We show our DNN-based PDF determination in the continuum limit, together with the 4-parameter fit results Model-$4p$, and the global analyses of the experimental data with NLO fixed order perturbation theory from xFitter~\cite{Novikov:2020snp} and JAM21nlo~\cite{Barry:2021osv} as well as the results considering threshold resummation the using double-Mellin method (JAM21nlonll).~\label{fig:PDFfinal}}
\end{figure}

We presented lattice calculations of the pion bi-local matrix elements for physical quark masses and a lattice spacing of $a=0.076$~fm. These results were combined with our previous results for $m_\pi=300$~MeV, but with very fine lattice spacing of $a=0.06$~fm and $a=0.04$~fm. This allowed us to obtain continuum-extrapolated results for pion valance PDF at the physical point. We used the NNLO short-distance factorization of RGI invariant matrix elements to determine up to $6^\mathrm{th}$ order Mellin moments of the pion valance PDF. The inclusion of the NNLO corrections to the perturbative matching did not result in significant changes to the numerical values of the moments, which indicates the convergence of leading-twist approximation with fixed-order perturbation theory. The NNLO correction also played a significant role in our test of the validity of the short-distance factorization as discussed in \app{dataevo}. The pion mass dependence of the moments were found to be very mild. As summarized in \fig{momscmp}, our results for the Mellin moments of the pion valance PDF are in excellent agreements with that from the different phenomenological global fits to experimental data. By fitting model functional forms of the PDF to the $zP_z$ dependence of RGI matrix elements we inferred its $x$ dependence. Further, we reconstructed the $x$ dependence of the PDF from the RGI matrix elements using a DNN. We found that the model fits and the DNN-based reconstruction of the PDF are in very good agreement among themselves, as well as with the phenomenological global fits of experimental data; see \fig{PDFfinal}. Next, we obtained the $x$-dependent quasi-PDF from the matrix elements, renormalized in a hybrid-scheme. From this quasi-PDF we determined the $x$ dependence of the PDF using NNLO perturbative matching. As illustrated in \fig{hybridPDFfinal}, we found that the pion mass dependence of the PDF to be small for pion momenta $\gtrsim1.5$~GeV and the $x$ dependence of the PDF are in good agreement with that from the DNN reconstruction and phenomenological global fit results. To conclude, we presented continuum-extrapolated lattice QCD results of the Mellin moments and the $x$ dependence of the valance PDF of pion for physcial values of quark masses. The Mellin moments and the $x$-dependent PDF, determined in multiple ways, agree with each other and with that form phenomenological global fits.

\section*{Acknowledgments}

This material is based upon work supported by: (i) The U.S. Department
of Energy, Office of Science, Office of Nuclear Physics through
Contract No.~DE-SC0012704; (ii) The U.S. Department of Energy,
Office of Science, Office of Nuclear Physics, from DE-AC02-06CH11357;
(iii) Jefferson Science Associates, LLC under U.S. DOE Contract
No.~DE-AC05-06OR23177 and in part by U.S. DOE grant No.~DE-FG02-04ER41302;
(iv) The U.S. Department of Energy, Office of Science, Office of
Nuclear Physics and Office of Advanced Scientific Computing Research
within the framework of Scientific Discovery through Advance Computing
(SciDAC) award Computing the Properties of Matter with Leadership
Computing Resources; (v) The U.S. Department of Energy, Office of
Science, Office of Nuclear Physics, within the framework of the TMD
Topical Collaboration. (vi) YZ is partially supported by an LDRD
initiative at Argonne National Laboratory under Project~No.~2020-0020.
(vii) SS is supported by the National Science Foundation under
CAREER Award PHY-1847893 and by the RHIC Physics Fellow Program of
the RIKEN BNL Research Center. 
(viii) KZ is supported by the BMBF under the ErUM-Data project and the AI grant of FIAS under SAMSON AG, Frankfurt.
(ix) S.Shi is supported by the U.S. Department of Energy, Office of Science, Office of Nuclear Physics, Grants Nos. DE-FG88ER41450 and DE-SC0012704.
(x) This research used awards of
computer time provided by the INCITE and ALCC programs at Oak Ridge
Leadership Computing Facility, a DOE Office of Science User Facility
operated under Contract No. DE-AC05-00OR22725. 
(xi) Computations
for this work were carried out in part on facilities of the USQCD
Collaboration, which are funded by the Office of Science of the
U.S. Department of Energy. (xii) Part of the data analysis are carried out on Swing, a high-performance computing cluster operated by the Laboratory Computing Resource Center at Argonne National Laboratory.

\appendix

\section{Truncation of the OPE formula}\label{app:momsTruncate}
\begin{figure}
\centering
	\includegraphics[width=0.4\textwidth]{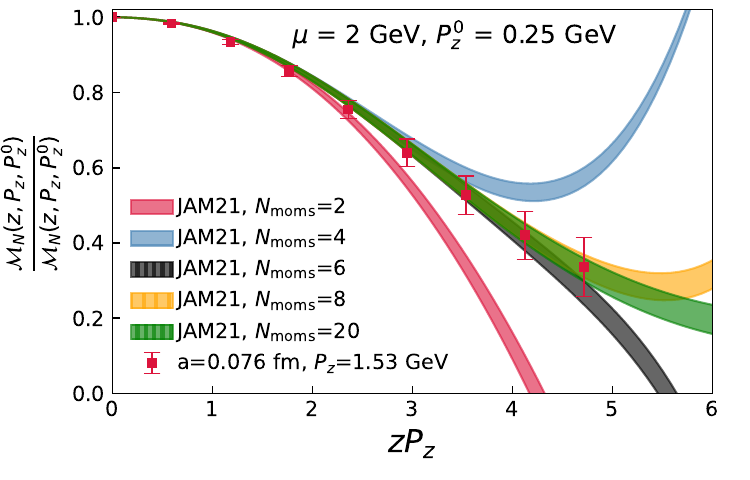}
	\includegraphics[width=0.4\textwidth]{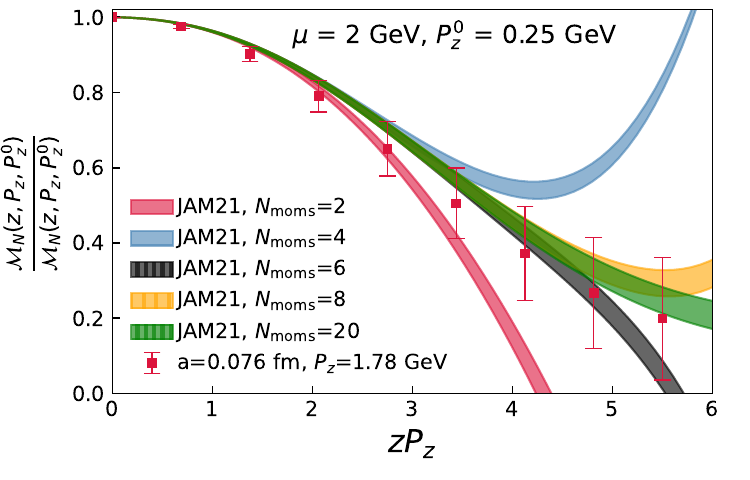}
	\caption{The reconstructed ratio-scheme matrix elements $\mathcal{M}(z,P_z,P_z^0)$ from the JAM21 global analysis \cite{Barry:2021osv} with $P_z^0$ = 0.25 GeV and NNLO matching are shown for $P_z=1.53$ GeV
	(top) and $P_z=1.78$ GeV (bottom). The width of the band corresponds
	to the uncertainty due to the variation of the renormalization scale. 
	\label{fig:momsTruncate}}
\end{figure}

In this appendix we discuss the truncation of the short distance factorization expression.
As mentioned in the main text, in lattice calculations one probes the rITD 
in a limited range of $z P_z$, namely up to $(zP_z)_{\rm{max}}\simeq 6$, and therefore, 
the short distance factorization formula
given by Eq. (\ref{eq:ratSDF}) can be truncated.
In order to test how many terms we need to keep in the sums entering Eq. (\ref{eq:ratSDF}) we construct
the rITD using the NNLO global analysis of the pion valence PDF by the JAM collaboration (JAM21) \cite{Barry:2021osv}
with NNLO matching, keeping different numbers of terms. We also compare it with our lattice results.
This analysis is shown in Fig. \ref{fig:momsTruncate}. Keeping 10 terms in the sum, i.e. considering up
to the 20th moment almost reproduces the exact result. Keeping only the first 4 terms in the sum results
in truncation errors that are smaller than the scale uncertainty, and given the errors of the lattice
results it is even possible to truncate the sum at the term proportional to $\langle x^6\rangle$. 
We note our lattice results for rITD agree very well with the one obtained from JAM21.

\section{The reliability of the perturbative matching at different $z$}\label{app:dataevo}

\begin{figure}
	\includegraphics[width=0.4\textwidth]{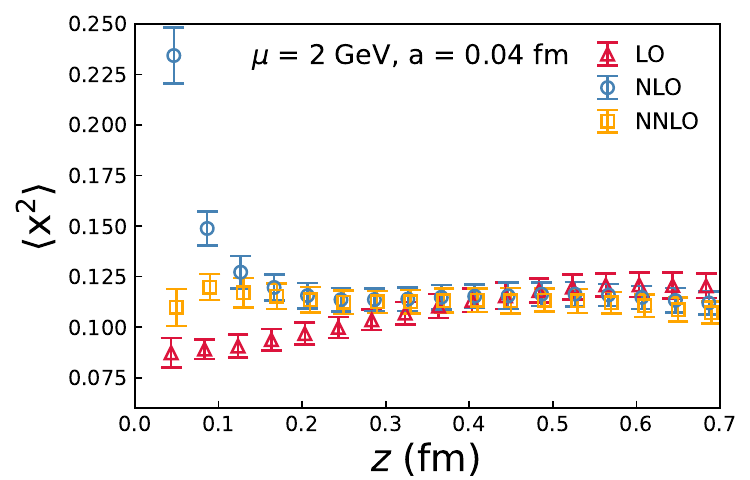}
	\includegraphics[width=0.4\textwidth]{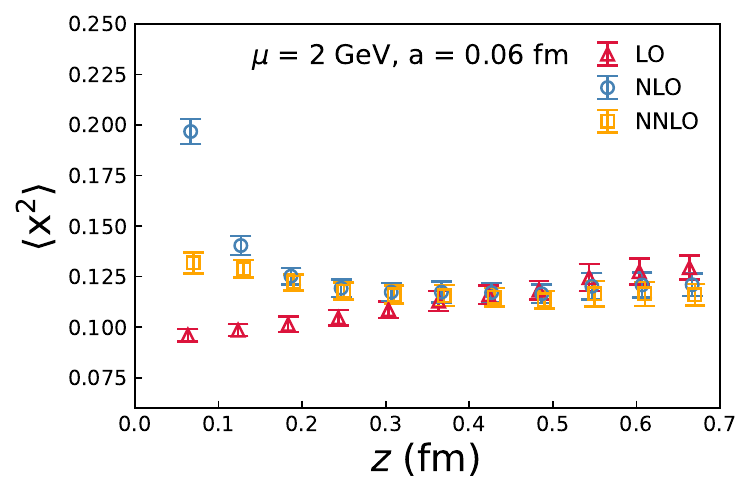}
	\caption{$\langle x^2 \rangle$ extracted from the ratio-scheme data $\mathcal{M}(z,P_z,P_z^0)$ at each $z$ using LO, NLO, NNLO kernels is shown for the a = 0.04 fm ensemble (top) and $0.06$ ensemble (bottom).\label{fig:momfine}}
\end{figure}

In this appendix we scrutinize the validity of the perturbative matching for different values 
of $z$. To do this we revisit our results for the ratio scheme matrix elements
and the extraction of the moments at smaller lattice spacings,
$a=0.04$ and $0.06$ fm, and unphysical pion mass, $m_{\pi}=300$ MeV. 
In Fig. \ref{fig:momfine} we show the second moment extracted for fixed values of $z$ at LO, NLO and NNLO.
We see the same tendencies as for $a=0.076$ fm shown in Fig. \ref{fig:momsingle076}. In particular, the NNLO
result has the smallest $z$ dependence. At the smallest two values of $z$ the second moment is too high.
However, in physical units
the values of $z$ for which the second moment is too high are shifted
to smaller $z$ confirming that this effect is mostly due to lattice discretization. 

The ratio-scheme matrix element defined in \Eq{ratioR} is an RGI quantity, from which we can extract the Mellin moments perturbatively through the leading-twist OPE formula~\Eq{ratSDF}. For the following discussion we consider the
$P_z=0$ case and write the OPE formula as 
\begin{align}\label{eq:ratioLO}
\begin{aligned}
\mathcal{M}(z,P_z,P_z^0=0)=\sum_{n=0}\frac{(-izP_z)^n}{n!}\langle x^n \rangle_{\rm LO}(z),
\end{aligned}
\end{align}
where we define
 \begin{align}\label{eq:momevo}
 	\langle x^n\rangle_{\rm LO}(z)=c_n(z^2\mu^2)\langle x^n\rangle(\mu),~c_n(z^2\mu^2)=\frac{C_n(z^2\mu^2)}{C_0(z^2\mu^2)}.
 \end{align}
On the lattice side $\langle x^n\rangle_{\rm LO}(z)$ is obtained by fitting the results for the rITD with the 
the polynomial form and the results are shown in Fig. \ref{fig:momsingle076} and Fig. \ref{fig:momfine} as open
triangles.
At leading order $c_n=1$ and 
$\langle x^n\rangle_{\rm LO}(z)$ is just the usual Mellin moment. 
Beyond leading order the $z$-dependence of $\langle x^n\rangle_{\rm LO}(z)$ should match the $z$-dependence
obtained in lattice QCD calculations if perturbation theory is sufficiently accurate. 
This ensures that the $\langle x^n\rangle(\mu)$ extracted from the lattice are independent of $z$.
We study $\langle x^n\rangle_{\rm LO}(z)$ for $x<0.3$ fm where perturbation 
theory may be reliable.

In this work, we used perturbative kernels up to NNLO level,
\begin{align}\label{eq:coeffNNLO}
    C_n^{\textup{NNLO}}(\mu^2z^2) = 1 + \frac{\alpha_s(\mu)}{2\pi} C^{(1)}_n(\mu^2z^2)+ \frac{\alpha_s^2(\mu)}{2\pi} C^{(2)}_n(\mu^2z^2),
\end{align}
where $C^{(1)}_n(\mu^2z^2)$ and $C^{(2)}_n(\mu^2z^2)$ are the 1-loop and 2-loop coefficients that contain terms proportional to $\textup{ln}(\mu^2z^2e^{2\gamma_E}/4)$.
For example, at NLO, $C^{(1)}_n(\mu^2z^2)$ reads~\cite{Izubuchi:2018srq},
 \begin{align}
 \begin{aligned}
 &C^{(1)}_n(z^2\mu^2)=C_F[(\frac{3+2n}{2+3n+n^2}+2H_n){\rm ln}\frac{\mu^2z^2e^{2\gamma_E}}{4}\\
 &+\frac{5+2n}{2+3n+n^2}+2(1-H_n)H_n-2H^{(2)}_n],
 \end{aligned}
 \end{align}
 with $H_n$ being the harmonic numbers.
If $\mu$ is not too different from $2/(ze^{\gamma_E})$ the logarithms are not too large and
the perturbative expansion for the Wilson coefficients is well behaved.
However, if $z$ is varied in a large range then the logarithms could become large
and need to be resummed. This can be done with the help of the RG equation
\begin{align}
\left(\frac{\partial}{\partial {\rm ln}\mu^2} + \beta(a_s)\frac{\partial}{\partial a_s} - \gamma_n \right) C_n(z^2\mu^2)=0,
\label{eq:ratioeq}
\end{align}
in which the anomalous dimension reads (up to NNLO),
\begin{align}
\begin{split}
\gamma_n
&= a_s [\frac{3C_F}{2}-\int_0^1dx x^n P^(0)_{qq}(x) ]\\
&+a_s^2\{\int_{0}^1dxx^n[P^{V(1)}_{qq}(x)\theta(x)-P^{(1)}_{q\overline{q}}(x)\theta(x)]\\
&+[C_F^2(-\frac{5}{8}+\frac{2\pi^2}{3})+C_FC_A(\frac{49}{24}-\frac{\pi^2}{6})+C_Fn_fT_F(-\frac{5}{6})]\}\\
&=a_s \gamma_n^{(1)}+a_s^2\gamma_n^{(2)} ,
\end{split}
\end{align}
and the running coupling $a_s=\alpha_s/(2\pi)$ obeys,
\begin{align}
\begin{aligned}
&\beta(a_s)=\frac{da_s}{d\textup{ln}\mu^2}=-a_s^2\beta_0-a_s^3\beta_1\,,\\
&\beta_0 = \frac{11C_A-4n_fT_F}{6}, \quad \beta_1 = \frac{102-38n_f/3}{4} .
\end{aligned}
\end{align}
Then solving the RG equation with boundary condition $Q_0=k\frac{2}{ze^{\gamma_E}}$, one obtains the leading-logarithm (LL) resummed NLO kernels (NLOevo),
\begin{align}\label{eq:coeffNLOevo}
\begin{aligned}
C_n^{\rm{NLO_{evo}}}(\mu^2z^2)=C_n^{\rm{NLO}}(Q_0^2z^2)(\frac{a_s(\mu)}{a_s(Q_0)})^{-\gamma_n^{(1)}/\beta_0}
\end{aligned}
\end{align}
or NLL resummed NNLO kernels (NNLOevo),
\begin{align}\label{eq:coeffNNLOevo}
\begin{aligned}
&C_n^{\rm{NNLO_{evo}}}(\mu^2z^2)=C_n^{\rm{NNLO}}(Q_0^2z^2)\\
&
\displaystyle
\times e^{-\frac{\gamma_n^{(1)}\textup{ln}\frac{a_s(\mu)}{a_s(Q_0)}}{\beta_0}-\frac{(-\beta_1\gamma_n^{(1)}+\beta_0\gamma_n^{(2)})\textup{ln}\frac{\beta_0+\beta_1a_s(\mu)}{\beta_0+\beta_1a_s(Q_0)}}{\beta_0\beta_1}}
\end{aligned}
\end{align}
where $\gamma_n^{(1)}$ and $\gamma_n^{(2)}$ are the anomalous dimension of the $n$th moments. For a conventional choice $k$ = 1 of $Q_0^2=\frac{4k^2}{z^2e^{2\gamma_E}}$, terms proportional to logarithms $\textup{ln}(\mu^2z^2e^{2\gamma_E}/4)$ are all cancelled. For example, the NLO + LL kernels read,
\begin{align}\label{eq:coeffNLOevo2}
\begin{aligned}
&C_n^{\rm{NLO_{evo}}}(\mu^2z^2)=\\
&\big\{1+\frac{\alpha_s(Q_0)}{2\pi}C_F\big[(\frac{3+2n}{2+3n+n^2}+2H_n)\ln(k^2)\\
&+\frac{5+2n}{2+3n+n^2}+2(1-H_n)H_n-2H^{(2)}_n\big]\big\}\left(\frac{a_s(\mu)}{a_s(Q_0)}\right)^{-\frac{\gamma_n^{(1)}}{\beta_0}} .
\end{aligned}
\end{align}

\begin{figure*}
\centering
	\includegraphics[width=0.4\textwidth]{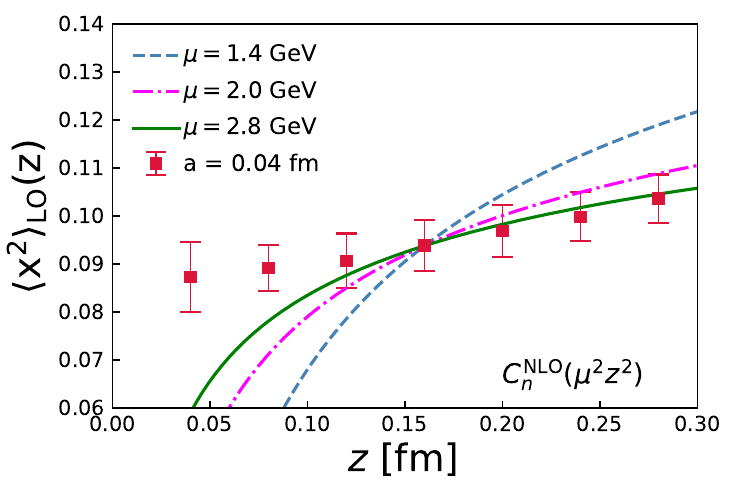}
	\includegraphics[width=0.4\textwidth]{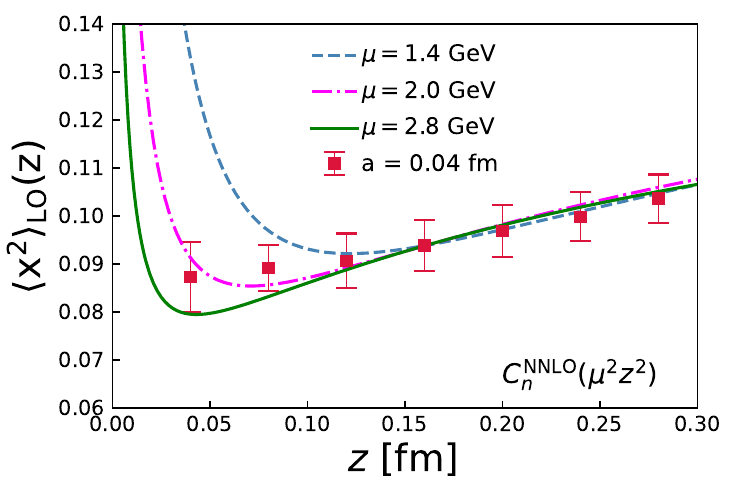}
	\includegraphics[width=0.4\textwidth]{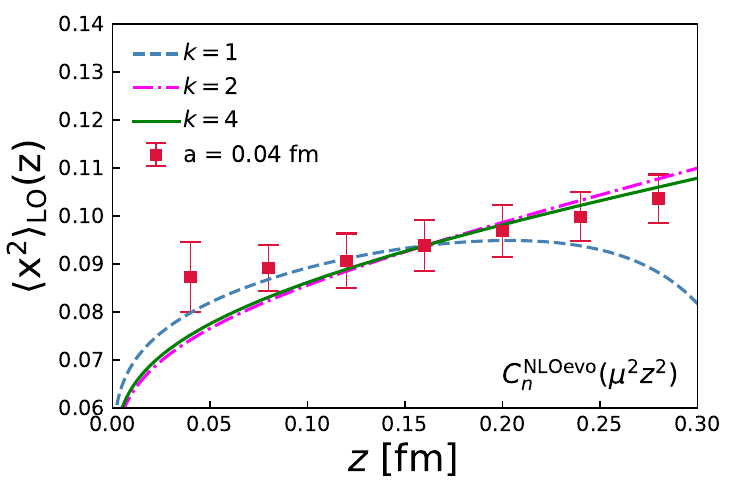}
	\includegraphics[width=0.4\textwidth]{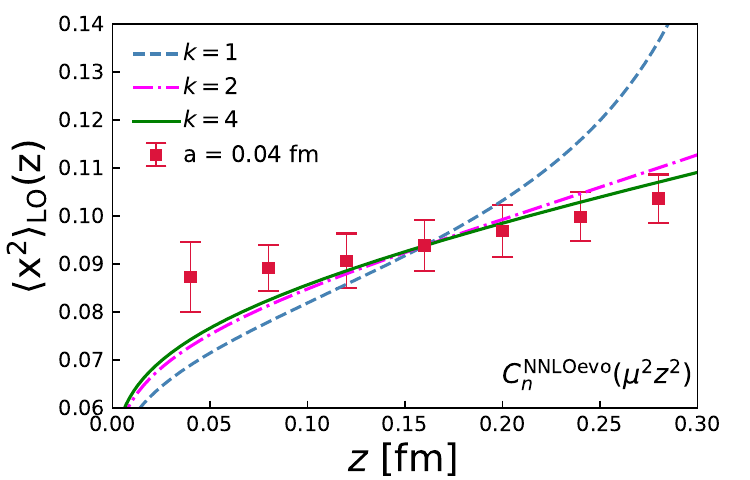}
	\caption{$\langle x^2 \rangle_{\rm LO}(z)$ extracted from the ratio
	scheme matrix elements for the $a=0.04$ fm ensemble compared to the perturbative
	predictions at NLO and NNLO (top panels),  and NLO$_{evo}$ and NNLO$_{evo}$ (bottom panels) for different renormalization scales.
	The perturbative results have been normalized to the lattice result
	at $z=0.16$ fm.
	\label{fig:momsevo}}
\end{figure*}

As discussed above, the coefficients $c_n(\mu^2z^2)$ are supposed to compensate the $z$ dependence of $\langle x^2 \rangle_{\rm LO}(z)$ and produce a $z$-independent $\langle x^2 \rangle(\mu)$. Now we can contrast the $z$-dependence
of $\langle x^2 \rangle_{\rm LO}(z)$ with the perturbative
prediction of $c_n(\mu^2 z^2)$ at different order and different choices of the renormalization scale.
We perform such a comparison for the $a=0.04$ fm ensemble,
where we have the largest range in $z$ and the smallest discretization errors.
To see the predictive power of the perturbative result, we can fix $\langle x^2 \rangle(\mu)$ at some $z_0$ then predict the $\langle x^2 \rangle_{\rm LO}(z)$ at different values of $z$. Here we choose $z_0$ = 0.16 fm which is in the middle of the range $z\in[0,0.3\rm~fm]$. In the upper panels of \fig{momsevo}, we show the fixed-order predictions, i.e. NLO (left) and NNLO (right), for $\langle x^2 \rangle_{\rm LO}(z)$. We choose the factorization scale $\mu=2$ GeV as the central value and vary it by a factor of $\sqrt{2}$. As one can see, the NLO result with $\mu\gtrsim$ 2 GeV can describe most of the data points.
For $z>0.1$ fm, the NLO result can describe the $z$ dependence of  
$\langle x^2 \rangle_{\rm LO}(z)$ reasonably well for $\mu=2$ GeV 
and $\mu=2.8$ GeV.
However, the  scale variation of the NLO result is quite large and
for lower values of the renormalization scale, e.g. $\mu$ = 1.4 GeV, the NLO result fails to describe the lattice data. 
On the contrary, the NNLO result
shows a very small scale dependence for $z>0.1$ fm and describes the lattice data very well. 
Thus the use of NNLO largely improves the quality of the
perturbative matching. For small separations, $z<0.1$ fm, the NLO result fails to describe the lattice calculations and has large scale dependence. The NNLO result for these distances does a better job in describing the lattice data but has large scale dependence. 
Furthermore, the NLO and NNLO results have very different shapes for $z<0.1$ fm,
explaining the discrepancy between the NLO and NNLO results seen in Fig. \ref{fig:momsingle076} and Fig.
\ref{fig:momfine} at the smallest $z$.
From this we conclude that there are large
logs in the perturbative result which need to be resummed.
The resummed results given by 
\Eq{coeffNLOevo} and \Eq{coeffNNLOevo}  are shown in the lower panels of \fig{momsevo}. 
Here we use $\mu=$ 2 GeV and set $k$ = 1, 2, 4 for $Q_0=k\frac{2}{ze^{\gamma_E}}$. As one can see,
the naive choice $Q_0=\frac{2}{ze^{\gamma_E}}$ ($k=1$) that eliminates all the logarithms $\textup{ln}(\mu^2z^2e^{2\gamma_E}/4)$ does not give the best
description of the lattice data points. When $k\gtrsim$ 2, the data under consideration can be well described except the first two points which suffer from significant discretization effects.
This suggests
$Q_0\simeq \frac{4}{ze^{\gamma_E}}$ is the better choice to set
the scale in $\alpha_s$ for the RG improved perturbative result.
We also note that NLO$_{evo}$ and NNLO$_{evo}$ results are very similar
and describe the lattice data for $z<0.3$ fm, except when the scale
$Q_0$ becomes too low and $\alpha_s$ is very large.

To summarize the above discussion: the fixed order NLO and NNLO results
describe the $z$-dependence of the lattice results well for $z>0.1$ fm
provided $\mu \ge 2$ GeV, but are not reliable for $z<0.1$ fm.
For NNLO, smaller values of the scale $\mu$ are possible since the scale variation
is tiny. The NNLO result also works at larger $z$ which shows a controlled scale
dependence.
The RG-improved results, NLO$_{evo}$ and NNLO$_{evo}$ describe
the lattice data well for $z<0.3$ fm, if the coupling is fixed
at $Q_0 \simeq \frac{4}{ze^{\gamma_E}}$, suggesting that this
scale sets the running of $\alpha_s$ in the perturbative
expression.

It may appear surprising that the NNLO result seems to
capture the $z$-dependence of the ratio scheme matrix elements 
at $z\simeq 0.3$ fm or even larger. In Ref. \cite{Gao:2020ito} it 
was suggested that this may be due to the fact that 
higher twist contributions, while non-negligible for $z>0.3$ fm,
cancel out in the ratio scheme. In Ref. \cite{Gao:2021dbh} it
was shown that the NNLO and NNNLO corrections to $C_0(\mu^2 z^2)$
are large for $z>0.3$ fm (c.f. Fig. 6 in that paper). Our analysis
suggest that some of these large corrections, which are also present in $C_n(\mu^2 z^2),~n>0$
cancel out in the ratio $c_n=C_n/C_0$, rendering the perturbative expansion reliable
even at relatively large values of $z$. The analysis in this appendix shows that
moments of the pion PDF can be reliably extracted using only $z<0.3$ fm if RG-improved matching is
used. In this region of $z$, the perturbative matching is reliable.
We also see, however, that using the fixed order NNLO result allows the moments to
be obtained reliably at large $z$, where the applicability of
the perturbative matching may seem to be doubtful. 
But the two determinations agree.
This is because some higher order perturbative corrections to the Wilson coefficients $C_n$ cancel out in $C_n/C_0$ and there is also cancellations of the higher
twist contributions.
This opens up the possibility to reliably determine the PDF using
the ratio scheme in current lattice calculations, which are typically performed on lattices with $a>0.04$ fm.

\section{Mass dependence of PDF from hybrid renormalization and $x$-space matching}\label{app:qPDFmass}
\begin{figure*}
\centering
	\includegraphics[width=0.32\textwidth]{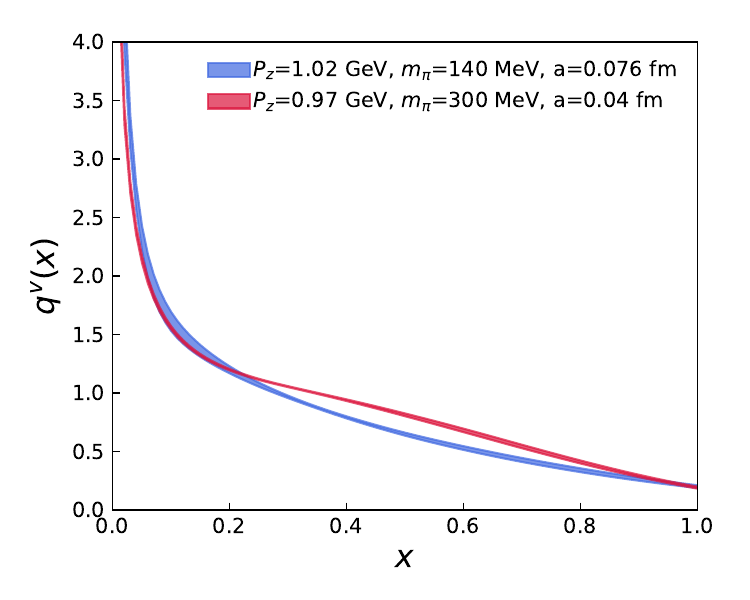}
	\includegraphics[width=0.32\textwidth]{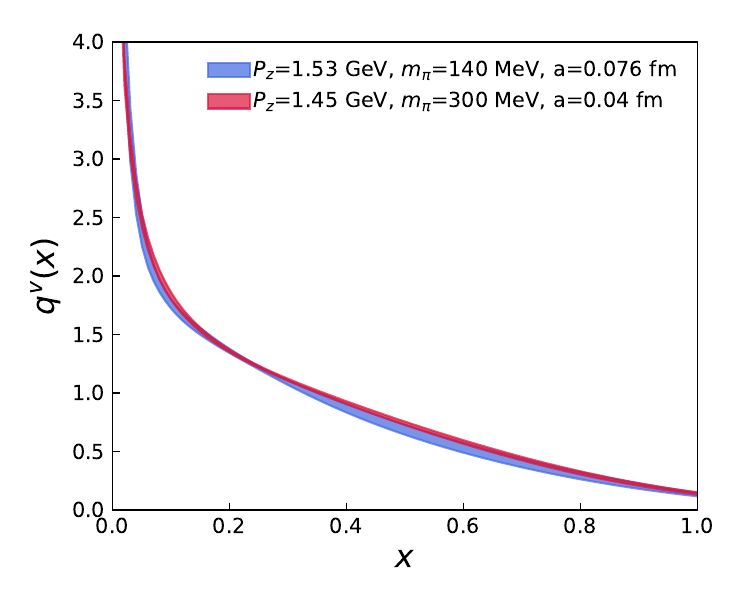}
	\includegraphics[width=0.32\textwidth]{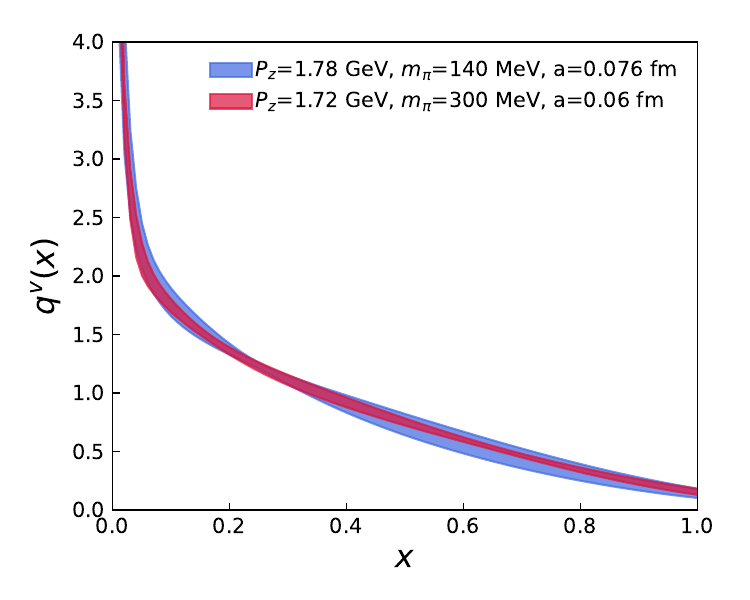}
	\caption{The pion valence distribution $q^v(x)$ obtained from an NNLO LaMET matching is shown. We show the results from the $m_\pi$ = 140 MeV case (blue) and the $m_\pi$ = 300 MeV cases~\cite{Gao:2021dbh} (red) for comparison.\label{fig:hybridPDFcmp}}
\end{figure*}

To understand the pion mass and lattice spacing dependence of
the results, we compare the pion valence PDF obtained
in this work with the earlier calculation performed
at two lattice spacings, $a=0.04$ fm and $a=0.06$ fm,
with a pion mass $m_{\pi}=300$ MeV \cite{Gao:2021dbh} in \fig{hybridPDFcmp}. It can be observed that the $q^v(x)$ from different calculations show some discrepancy at $P_z\sim1$ GeV, while they start to overlap when $P_z\gtrsim1.5$ GeV. On the one hand the lattice spacing dependence is mild as expected from previous sections, on the other hand it also suggests the pion mass may play a role for low momenta but decays rapidly for large $P_z$.

\section{The matching strategy and the DNN representation of the ITD $Q(\lambda,\mu)$ }\label{app:DNNsetup}
In \sec{DNNITD} we apply \Eq{OPEtw2ITD} to solve for the light-cone ITD $Q(\lambda,\mu)$ from the ratio-scheme matrix elements. Starting from the standard reduced Ioffe-time distribution (rITD) one has the factorization formula,
\begin{align}
\begin{split}
&\widetilde{Q}(z,P_z)\\
=&\int_{-1}^1 d\alpha \frac{\mathcal{C}(\alpha,\mu^2z^2)}{C_0^{\overline{\rm{MS}}}(\mu^2z^2)} Q(\alpha\lambda,\mu) \\
=&Q(\lambda,\mu) + \int_{-1}^1 d\alpha \big[Q(\alpha \lambda,\mu)-Q(\lambda,\mu)\big] \frac{n_\text{NNLO}(\alpha, \mu^2 z^2)}{d_\text{NNLO}(\mu^2z^2)},
\end{split}
\end{align}
where we re-write the perturbative kernels $\overline{\mathcal{C}}(\alpha,\mu^2z^2)$ by,
\begin{align}
\begin{split}
n_\text{NNLO}(\alpha, \mu^2 z^2) 
&=\sum_{i=1}^{2}\sum_{j=0}^{i} \Big({\alpha_s \over 2\pi}\Big)^in_{i,j}(\alpha) \Big[L(\mu^2 z^2)\Big]^j,\\
d_\text{NNLO}(\mu^2 z^2)
&=1 + \sum_{i=1}^{2}\sum_{j=0}^{i} \Big({\alpha_s \over 2\pi}\Big)^i d_{i,j} \Big[L(\mu^2 z^2)\Big]^j,
\end{split}
\end{align}
and,
\begin{align}
&L(\mu^2 z^2) \equiv 2\gamma_E + \ln \frac{\mu^2 z^2}{4},\\
&d_{1,0}=\frac{5C_F}{2},\\
&d_{1,1}=\frac{3C_F}{2},\\
\begin{split}
&d_{2,0}=C_F\Bigg[\frac{469\beta_0}{48}+\frac{223C_F-94C_A}{96}\\
&+\frac{\pi^2 (8C_F-15C_A)}{36}+2(C_A-4C_F)\zeta(3)\Bigg],
\end{split}\\
\begin{split}
&d_{2,1}=\Bigg[C_F^2\left(-\frac{5}{8}+\frac{2\pi^2}{3}\right)+C_FC_A\left(\frac{49}{24}-\frac{\pi^2}{6}\right) \\
&-\frac{5}{6}C_Fn_fT_F \Bigg]+ \frac{5 C_F}{2}\Big(\frac{3}{2}C_F + \beta_0 \Big),\\
\end{split}\\
&d_{2,2}=\frac{3C_F}{4}\Big(\frac{3}{2}C_F + \beta_0 \Big),\\
&n_{1,0}(\alpha)=C_F\Big[2(1-\alpha)-\frac{1+\alpha^2}{1-\alpha}-\frac{4\ln(1-\alpha)}{1-\alpha}\Big],\\
&n_{1,1}(\alpha)=-C_F\frac{1+\alpha^2}{1-\alpha} ,
\end{align}
as well as $n_{2,0}(\alpha)$, $n_{2,1}(\alpha)$ and $n_{2,2}(\alpha)$ in more complicated forms. The constants in the formulas are $C_F=4/3$, $T_F=1/2$, $C_A=3$, $n_f=3$ (3 flavor in this work) and $\beta_0=(11C_A-4n_fT_F)/6$. With this setup, we then numerically saved $n_\text{NNLO}(\alpha, \mu^2 z^2)$ (mainly for $n_{2,j}(\alpha)$) for frequent calls.

As discussed in the main text of \sec{DNNITD}, we express the light-cone ITD $Q(\lambda,\mu)$ by the deep neural network (DNN),
\begin{align}
    Q_\textup{DNN}(\lambda, \mu) \equiv  \frac{f_\textup{DNN}(\boldsymbol{\theta};\lambda)}{f_\textup{DNN}(\boldsymbol{\theta};0)},
\end{align}
and minimize the loss function,
\begin{align}
\begin{split}
J(\boldsymbol{\theta},r_\text{mod}) 
\equiv
    \frac{\eta}{2} \boldsymbol{\theta}\cdot\boldsymbol{\theta}+\frac{1}{2} \chi^2(\boldsymbol{\theta},r_\text{mod}) .
\end{split}
\end{align}
The $\frac{\eta}{2} \boldsymbol{\theta}\cdot\boldsymbol{\theta}$ term is to make sure the DNN represented function has good shape and is smooth. The $\chi^2$ is defined as,
\begin{align}
\begin{split}
&\chi^2(\boldsymbol{\theta},r_\text{mod})\\
&=\sum_{P_z\textgreater P_z^0}^{P_z^{\textup{max}}}\sum^{z_\textup{max}}_{z_\textup{min}}\frac{(\mathcal{M}(z,P_z,P_z^0)-\mathcal{M}_{\textup{DNN}}(z,P_z,P_z^0;\boldsymbol{\theta},r_\text{mod}))^2}{\sigma^2(z,P_z,P_z^0)} ,
\end{split}
\end{align}
with $\sigma(z,P_z,P_z^0)$ being the statistical errors of the ratio-scheme matrix elements $\mathcal{M}(z,P_z,P_z^0)$. We analytically solve the gradients of $\chi^2$,
\begin{align}
\begin{split}
&\frac{\partial \chi^2}{2\partial r_\text{mod}}\\
=&\sum_{P_z\textgreater P_z^0}^{P_z^{\textup{max}}}\sum^{z_\textup{max}}_{z_\textup{min}}  \frac{\mathcal{M}(z,P_z,P_z^0)-\mathcal{M}_{\text{DNN}}(z,P_z,P_z^0)}{\sigma^2(z,P_z,P_z^0)[\widetilde{Q}_\text{DNN}(z,P_z^0)+r_{\textup{mod}}(aP_z^0)^2]}\\
\times &\Big[(aP_z)^2 - \mathcal{M}_{\text{DNN}}(z,P_z,P_z^0)(aP_z^0)^2 \Big],\\[1mm]
\end{split}
\end{align}
\begin{align}
\begin{split}
&\frac{\nabla_{\boldsymbol{\theta}} \chi^2}{2}\\
&=\sum_{P_z\textgreater P_z^0}^{P_z^{\textup{max}}}\sum^{z_\textup{max}}_{z_\textup{min}}  \frac{\mathcal{M}(z,P_z,P_z^0)-\mathcal{M}_{\text{DNN}}(z,P_z,P_z^0)}{\sigma^2(z,P_z,P_z^0)[\widetilde{Q}_\text{DNN}(z,P_z^0)+r_{\textup{mod}}(aP_z^0)^2]}\\
&\times\frac{1}{f_\text{DNN}(0)}\Bigg\{
    \nabla_{\boldsymbol{\theta}} f_\text{DNN}(zP_z)\\
 &+\int_{-1}^{1}
    \big[\nabla_{\boldsymbol\theta} f_{\text{DNN}}(\alpha zP_z)-
    \nabla_{\boldsymbol\theta} f_{\text{DNN}}(zP_z)\big]\,
    \frac{n_\text{NNLO}(\alpha, \mu^2 z^2)}{d_\text{NNLO}(\mu^2 z^2)}  d\alpha\\
    &-\mathcal{M}_{\text{DNN}}(z,P_z,P_z^0)
    \Big[\nabla_{\boldsymbol{\theta}} f_\text{DNN}(P_z^0 z)
    \\
    &+ \int_{-1}^{1}\!\!
    \big[\nabla_{\boldsymbol\theta} f_{\text{DNN}}(\alpha P_z^0 z)-
    \nabla_{\boldsymbol\theta} f_{\text{DNN}}(P_z^0 z)\big]\,
    \frac{n_\text{NNLO}(\alpha, \mu^2 z^2)}{d_\text{NNLO}(\mu^2 z^2)}  d\alpha
    \Big]\\
    &+r_\text{mod} [(aP_z)^2-\mathcal{M}_{\text{DNN}}(z,P_z,P_z^0) (aP_z^0)^2] \nabla_{\boldsymbol\theta} f_{\text{DNN}}(0)
    \Bigg\} .
\end{split}
\end{align}
\vskip0.7truecm
To optimize the parameters, we apply the Adam optimization method~\cite{kingma2014adam} for the gradient descent in this work.

\newpage

\bibliographystyle{apsrev4-1.bst}
\bibliography{ref}

\end{document}